\documentclass[a4paper,fleqn,usenatbib]{mnras}

\usepackage{newtxtext,newtxmath}
\usepackage[T1]{fontenc}
\usepackage{ae,aecompl}

\usepackage{graphicx}	% Including figure files
\usepackage{amsmath}	% Advanced maths commands
\usepackage{amssymb}	% Extra maths symbols
\usepackage{xcolor, comment, ulem, epstopdf, mathtools, amsfonts, float, color}

\newcommand{\rmd}{\,\mathrm{d}}
\newcommand{\rmi}{\,\mathrm{i}}
\newcommand{\del}{\partial}

\title[Magneto-gravity rays in evolved stellar cores]{Magneto-gravity wave packet dynamics in strongly magnetised cores of evolved stars}

\author[S. T. Loi]{
Shyeh Tjing Loi\thanks{E-mail: stl36@cam.ac.uk}
\\
Department of Applied Mathematics and Theoretical Physics, University of Cambridge, Centre for Mathematical Sciences, Wilberforce Road, Cambridge CB3 0WA, UK
}

\date{Accepted XXX. Received YYY; in original form ZZZ}

\pubyear{2019}

\begin{document}
\label{firstpage}
\pagerange{\pageref{firstpage}--\pageref{lastpage}}
\maketitle

\begin{abstract}
  Magnetic fields are believed to be generated in the cores of massive main sequence stars, and these may survive on to later stages of evolution. Observations of depressed dipole modes in red giant stars have been touted as evidence for these fields, but the predictions of existing magnetic theories have difficulty accommodating several aspects, including the need to return a fraction of wave energy from the core to the envelope, and the persistent gravity-like character of affected modes. In this work we perform a Hamiltonian ray tracing study investigating the dynamics of magneto-gravity waves in full spherical geometry, using realistic stellar models and magnetic field configurations. This technique applies in the limit where wavelengths are much shorter than scales of background variation. We conduct a comprehensive exploration of parameter space, examining the roles of wave frequency, spherical harmonic degree, wavevector polarisation, incoming latitude, field strength, field radius, and evolutionary state. We demonstrate that even in the presence of a strong field, there exist trajectories where waves remain predominantly gravity-like in character, and these are able to undergo reflection out of the core much like pure gravity waves. The remaining trajectories are ones where waves acquire significant Alfv\'{e}n character, becoming trapped and eventually dissipated. Orientation effects, i.e.~wavevector polarisation and incoming latitude, are found to be crucial factors in determining the outcome (trapped versus reflected) of individual wave packets. The allowance for partial energy return from the core offers a solution to the conundrum faced by the magnetic hypothesis.
\end{abstract}

\begin{keywords}
  MHD -- methods: numerical -- stars: interiors -- stars: magnetic field -- waves
\end{keywords}

%%%%%%%%%%%%%%%%%%%%%%%%%%%%%%%%%%%%%%%%%%%%%%%%%%

%%%%%%%%%%%%%%%%% BODY OF PAPER %%%%%%%%%%%%%%%%%%

%% INTRODUCTION
\section{Introduction}
Convection is believed to be a key ingredient in dynamo operation, the process by which magnetic fields are generated in conducting fluids \citep{Mestel2012}. Many stars are convective in some part of their volume; on the main sequence, stars more massive than about 1.2\,M$_\odot$ have convective cores, while those of masses between about 0.35\,M$_\odot$ and 1.2\,M$_\odot$ have convective envelopes, and stars of lower masses still are fully convective \citep{Maeder2008}. Evidence of surface magnetism, such as starspots and high-energy radiation associated with stellar flares, is widely observed in those stars whose convective zones extend to the surface \citep{Donati2009}. In contrast, stars more massive than 1.2\,M$_\odot$ show much weaker surface magnetic activity. However, the possibility of a dynamo operating in their cores has received interest over the years \citep{Krause1976}, the primary means of investigating this being through numerical simulations \citep{Brun2005, Featherstone2009}. It is thought unlikely that such fields would be able to reach the surface within a main-sequence lifetime, where they might be directly observed \citep{MacDonald2004}.

An unanticipated and controversial potential signature of deep interior magnetism, first reported by \citet{Mosser2012}, is a dichotomy of dipole mode amplitudes in asteroseismic observations of evolved stars. It has been established that in about 20\% of the red giant population, dipole modes exhibit amplitudes several times lower than in the remainder of the population, and furthermore the phenomenon is restricted to those stars massive enough to previously have had convective cores when on the main sequence \citep{Stello2016}. Affected stars have been termed ``depressed'', while the remainder are considered ``normal''. Follow-up analyses of the red giant data showed that higher multipole (quadrupole and octupole) modes also exhibit a dichotomy in their amplitudes, while radial modes appear not to be affected \citep{Stello2016a}. Out of the two main types of wave that can propagate in stars (acoustic and gravity), it is gravity waves that only exist for non-radial degrees, while no such restriction applies to acoustic waves. This suggests the crucial role of gravity waves, which are known to be localised to the core. Thus the observations generally point to the dichotomy being due to a difference in core properties between the two groups of stars. The restriction to stars previously able to host a convective core dynamo further draws a link to the possible role of deep interior magnetism, a suggestion first made by \citet{Garcia2014} in the course of a case study.

The inclusion of magnetic fields makes the treatment of stellar oscillations more difficult, since they necessarily break the spherical symmetry, increasing the dimensionality of the problem. Stable realistic magnetic field configurations one might find in nature are also thought to be highly non-trivial, resembling twisted tori with comparable poloidal and toroidal components \citep{Markey1973, Flowers1977, Braithwaite2004}. In spite of these complications, one technique that remains efficient and useful for studying wave propagation is Hamiltonian ray tracing, which is valid in the limit where wavelengths are much smaller than the length scales of background variation. This has had successful application in many problems of wave/continuum mechanics where the above assumption holds, including EM wave propagation through the Earth's ionosphere and magnetosphere \citep{Haselgrove1955, Walker2004, Fung2005}, seismic waves in the Earth's interior \citep{Cerveny1979, Chian1994}, and gravity waves in the Earth's atmosphere and ocean \citep{Broutman2004, Hasha2008}. In more astrophysical contexts, ray tracing forms the basis of time-distance helioseismology \citep{DuvallJr1993, DSilva1996}, and has been used to study magneto-acoustic wave propagation in the solar envelope \citep{Cally2006, Cally2006a}. It is being increasingly used to tackle problems in asteroseismology where spherical symmetry is significantly broken, for example inertial-acoustic \citep{Lignieres2008, Lignieres2009, Lignieres2010, Pasek2012}, gravito-inertial \citep{Prat2016, Prat2017, Prat2018} and magneto-gravito-inertial \citep{Valade2018} wave propagation in rapidly rotating stars.

Magneto-gravity ray tracing was invoked by \citet{Fuller2015} in a first attempt to connect magnetic fields to the red giant dipole dichotomy problem, wherein they suggested that gravity waves propagating into a region of sufficiently strong magnetic field would undergo scattering to higher wavenumbers and be dissipated. They predicted a 100\% loss of energy from the portion of the mode overlapping with the core (the g-like portion), resulting in a mode that is purely p-like in character. However, further analyses of the data by \citet{Mosser2017a} showed that the g-like character of affected modes is in general retained, and furthermore the measured mode amplitudes are inconsistent with 100\% energy loss within the core. A separate study by \citet{Arentoft2017} reached similar conclusions. Preliminary theoretical work by \citet{Loi2017} invoking resonant interactions of global modes with Alfv\'{e}n waves appeared to yield a mechanism that could produce partial damping, but difficulties lay in the self-consistency of input assumptions. Subsequent work by \citet{Loi2018} (hereafter LP18) studying the basic local interactions of gravity waves with strong magnetic fields in a simple Cartesian setup found that even if field strengths exceed the threshold quoted by \citet{Fuller2015}, it is not necessarily the case that all ingoing waves get trapped and dissipated. Rather, some may end up being reflected in a near-specular fashion, the outcome depending on the relative orientations of the wavevector, magnetic field lines and direction of stratification.

Taken in the context of the red giant dichotomy problem, the results of LP18 hint that orientation effects, if properly accounted for, can potentially enable the magnetic field hypothesis to be reconciled with observations. However, the problem setup was highly idealised, this being within a periodic Cartesian box bearing little resemblance geometrically to an actual star. Since geometry is closely linked to orientation, this aspect warrants further investigation which is the objective here. In this work, we undertake an exploration of parameter space, using realistic evolved stellar models and magnetic field configurations, to measure the fraction of ingoing gravity waves trapped/reflected by a magnetic field embedded in the core of a star, as a function of five parameters. These are the evolutionary state, field strength, radial extent of the magnetised region, wave frequency and spherical degree. To quantify these fractions we use a Hamiltonian ray tracing approach that involves launching a large number of rays of varying angular position and wavevector orientation into the core, and then tallying how many rays underwent trapping or reflection, for each combination of parameters. Trapped waves are known to diverge in wavenumber, implying that under conditions of non-zero viscosity/resistivity they will eventually dissipate, representing a loss route of energy which can lead to damping of a global mode of oscillation. By averaging over the surface of the star and directions of propagation, we then compute the trapping fraction experienced by a global mode, i.e.~how much energy tunnelling into the core gets dissipated, thereby linking our results to the broader problem.

This paper is organised as follows. In Section \ref{sec:models} we introduce the stellar models, magnetic field configurations and parameter space tested. In Section \ref{sec:methods} we derive the relevant equations for magneto-gravity ray tracing and describe the procedures for initialising, integrating, analysing and classifying the rays. In Section \ref{sec:results} we present the results for trapped and reflected fractions as a function of the five parameters listed above, then discuss their broader relevance, limitations and further implications in Section \ref{sec:discussion}. Finally, we conclude in Section \ref{sec:summary}.

%% MODELS & PARAMETERS
\section{Models and Parameters}\label{sec:models}
\subsection{Stellar models}
\subsubsection{Generation}
We generated a series of 2\,M$_\odot$ models along a single evolutionary track using the publicly-available stellar evolutionary code `Modules for Experiments in Stellar Evolution' (MESA, version r11701). Evolution was terminated artificially at an age of 1.076 Gyr, just before helium ignition. The Hertzsprung-Russell diagram for this stellar model is shown in Figure \ref{fig:2Msun_HR}. The full MESA inlist is contained in Appendix \ref{sec:inlist}. 

For the purposes of this work, three snapshots were selected. These correspond to a subgiant, young red giant and older red giant, hereafter referred to as Models A, B and C, and are marked in Figure \ref{fig:2Msun_HR}. Their ages, radii $R_*$, central pressures $p_c$, central densities $\rho_c$, dynamical speeds
\begin{align}
  V_\text{dyn} \equiv \sqrt{\frac{GM_*}{R_*}} \:,
\end{align}
and dynamical frequencies (large separations) $\nu_\text{dyn} = \omega_\text{dyn}/2\pi$, where
\begin{align}
  \omega_\text{dyn} \equiv \sqrt{\frac{GM_*}{R_*^3}}
\end{align}
and $M_*$ is the stellar mass, are listed in Table \ref{tab:profiles}.

\begin{table}
  \centering
  \caption{Parameters for the three snapshots indicated in Figure \ref{fig:2Msun_HR}, for which $M_* = 2$\,M$_\odot$.}
  \label{tab:profiles}
  \begin{tabular}{rccc}
    \hline
    Model & A & B & C \\ \hline
    Age (Gyr) & 0.9757 & 1.001 & 1.046 \\
    $R_*$ (R$_\odot$) & 4.3 & 6.1 & 15 \\
    $p_c$ (Pa) & $8.9 \times 10^{17}$ & $2.7 \times 10^{19}$ & $2.3 \times 10^{20}$ \\
    $\rho_c$ (kg/m$^3$) & $4.2 \times 10^6$ & $4.6 \times 10^7$ & $2.0 \times 10^8$ \\
    $V_\text{dyn}$ (m/s) & $3.0 \times 10^5$ & $2.5 \times 10^5$ & $1.6 \times 10^5$ \\
    $\nu_\text{dyn}$ ($\upmu$Hz) & 16 & 9.3 & 2.4 \\
    $V_{A,\text{crit}}$ (m/s) & $10^3$ & 50 & 1 \\
    $B_\text{crit}$ (MG) & 30 & 4 & 0.2 \\
    \hline
  \end{tabular}
\end{table}

\begin{figure}
  \centering
  \includegraphics[width=\columnwidth]{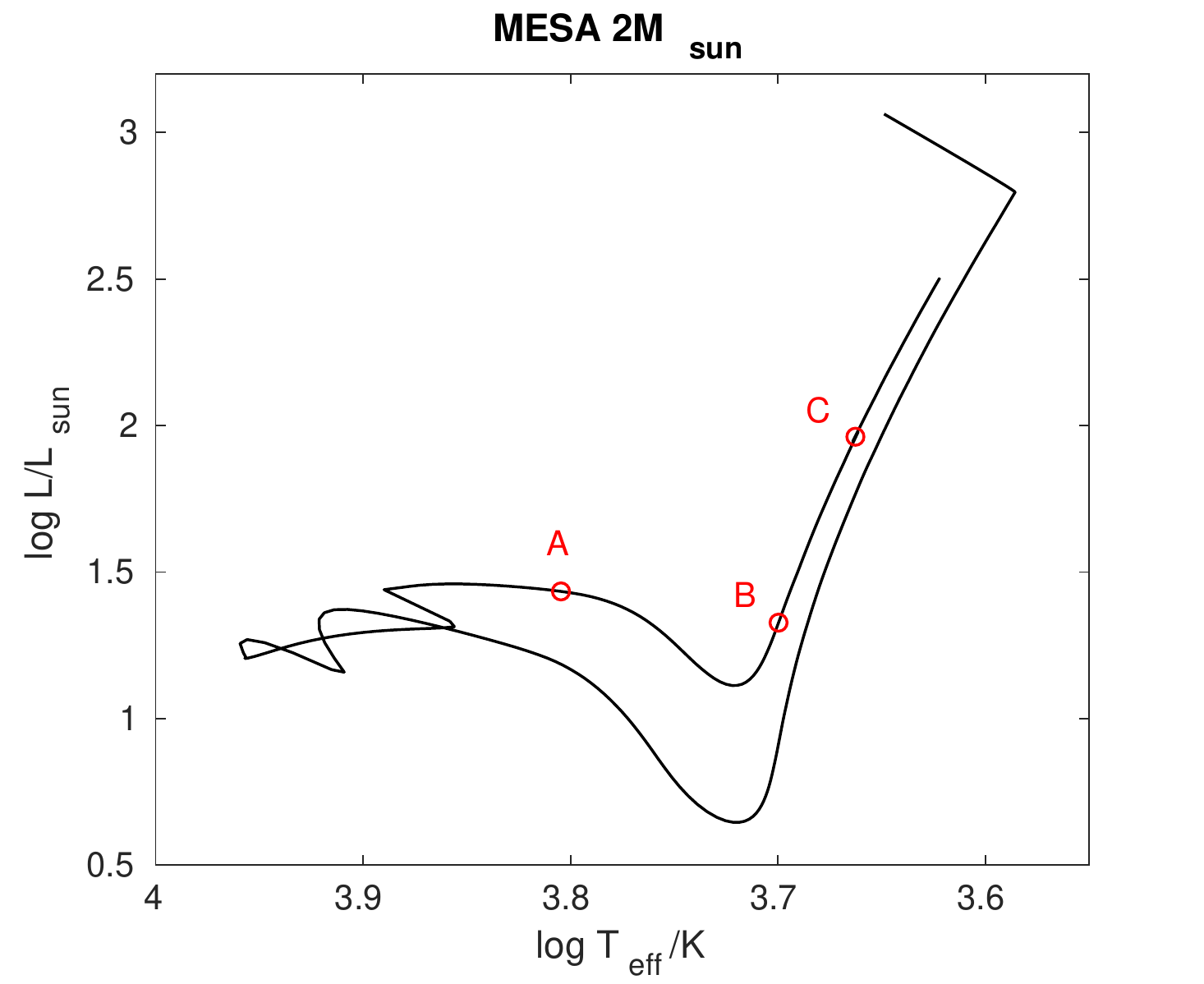}
  \caption{Hertzsprung-Russell diagram of the 2\,M$_\odot$ evolutionary sequence generated using MESA, showing the locations (red circles) of the three snapshots used in this work. They correspond to a subgiant star (Model A), a young RGB star (Model B), and an older RGB star (Model C), with respective ages of 975.7\,Myr, 1.001\,Gyr and 1.046\,Gyr.}
  \label{fig:2Msun_HR}
\end{figure}

\subsubsection{Post-processing}
It was noticed that some coarseness on the grid scale was generally present in the MESA output. Attempts were made to remedy this by tightening various mesh convergence criteria (see inlist in Appendix \ref{sec:inlist}). In addition, we smoothed the relevant output quantities (namely the density $\rho$, pressure $p$, gravitational field $g$ and adiabatic index $\Gamma_1$) by replacing each point by that lying on the least-squares line of best fit through the neighbouring $n$ points. We chose a window size of $n = 13$. The quantities were then interpolated to uniform radial grids of 1043, 1928 and 7693 points for Models A, B and C (the original non-uniform, centrally-condensed grids had 5217, 6428 and 7693 points, respectively), in a further effort to remove grid-scale coarseness in the core region. Note that the chosen mesh criteria meant that MESA was forced to use a much finer grid than normal, which allowed for scope for downsampling that would still retain large-scale features while eliminating grid-scale ones. Derivatives were approximated by central differencing, i.e.~$(q_{i+1}-q_{i-1})/(2h)$ for the derivative of the quantity $q$ at gridpoint $i$, where $h$ is the uniform grid spacing. The buoyancy frequency $N$ was computed via
\begin{align}
  N^2 \equiv g \left( \frac{1}{\Gamma_1 p} \frac{\rmd p}{\rmd r} - \frac{1}{\rho} \frac{\rmd \rho}{\rmd r} \right) \:,
\end{align}
and is shown in Figure \ref{fig:N_profiles_field_radii} for the three models. For subsequent ray tracing, the resultant grids of relevant quantities ($\rho$, $\rmd \rho/\rmd r$, $N$ and $\rmd N/\rmd r$) formed the basis of lookup tables, with intermediate values obtained through linear interpolation between neighbouring points.

In all plots and computations, it is the non-dimensional versions of all quantities that are used: frequencies are given in terms of $\omega_\text{dyn}$, speeds/velocities in terms of $V_\text{dyn}$, densities in terms of $\rho_c$, length scales in terms of $R_*$ and time scales in terms of $1/\omega_\text{dyn}$. Besides avoiding the usual numerical underflow/overflow problems, it also allows for meaningful comparison of certain quantities between the three models, which have very different $R_*$, $p_c$ and $\rho_c$. Table \ref{tab:profiles} lists the dimensional values of the various scaling factors.

\subsection{Magnetic field}
\subsubsection{Choice of field radii}\label{sec:Rf}
\begin{figure}
  \centering
  \includegraphics[width=\columnwidth]{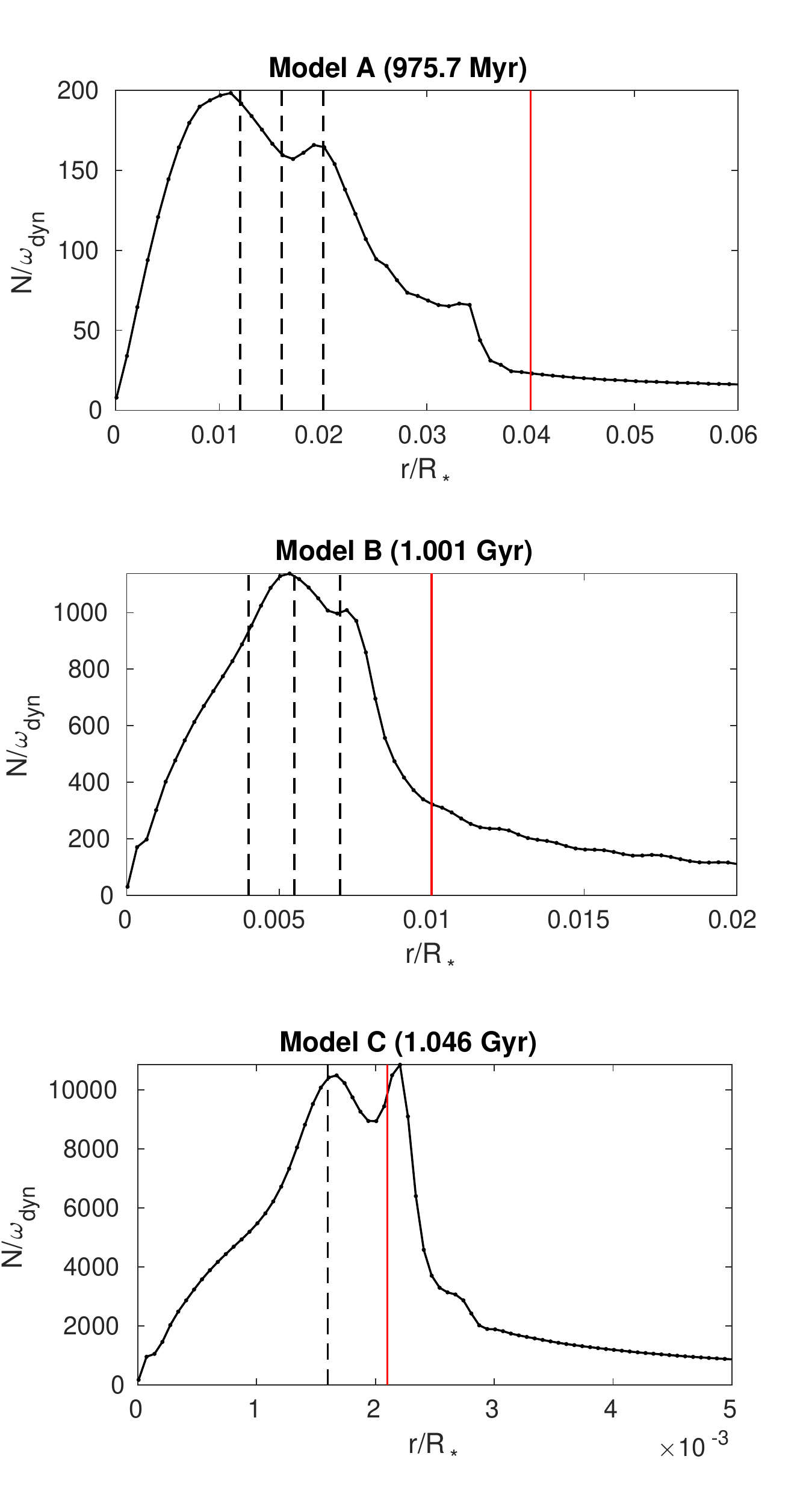}
  \caption{Profiles of the buoyancy frequency (in units of $\omega_\text{dyn}$) for each of the three evolutionary snapshots, as a function of radial distance. Note that the plots are zoomed in to the core region. Vertical black dashed lines indicate the radii chosen for generating magnetic field solutions, while the solid red lines mark the radial coordinate $r_0$ from which rays were launched into the core. For the middle panel, the form of the magnetic field for the three solutions is shown in Figure \ref{fig:Prendergast_profile30}.}
  \label{fig:N_profiles_field_radii}
\end{figure}

If the magnetic field had been generated by a previous core dynamo, it would be expected to occupy the region within the H-burning shell, which marks the boundary of the original convective core. We identified this location as the spike in $N$ just beyond the main maximum, corresponding to steep composition gradients where the H mass fraction sharply rose from zero (this was seen to occur near $r \approx 0.02\,R_*$, $0.0075\,R_*$ and $2.1 \times 10^{-3}\,R_*$ for Models A, B and C respectively). We imposed magnetic fields within the central $r = R_f$ of each stellar model, where the largest values of $R_f$ were in the vicinity of the H-burning shell, and we allowed $R_f$ to vary by nearly a factor of two. The values of $R_f$ selected for testing are shown in Figure \ref{fig:N_profiles_field_radii} as dashed vertical lines: for Model A,
\begin{align}
  R_f &\in \{ 0.012, 0.016, 0.02 \}\,R_* \:, \\
  \shortintertext{while for Model B,}
  R_f &\in \{ 4, 5.5, 7 \} \times 10^{-3}\,R_* \:,
\end{align}
and for Model C, $R_f = 1.6 \times 10^{-3}\,R_*$.

\subsubsection{Magnetic field configuration}
It has been established through analytical and numerical means \citep{Tayler1973, Flowers1977, Braithwaite2006} that stable magnetic equilibria necessarily involve a mixture of poloidal and toroidal components, and it is thought that such configurations are likely to take the form of twisted tori \citep{Yoshida2006, Braithwaite2017}. An analytic solution describing an axisymmetric twisted torus equilibrium field confined to a spherical volume was first derived by \citet{Prendergast1956} for incompressible fluids, and later extended to the compressible case by \citet{Duez2010a} and \citet{Duez2010}. It is the general (compressible) solution that we implement in this work, whose derivation we now briefly overview.

One begins by imposing the force-balance equation (note that the effects of rotation are neglected)
\begin{align}
  \nabla p + \rho \nabla \Phi = \frac{1}{\mu_0} (\nabla \times \mathbf{B}) \times \mathbf{B} \:, \label{eq:force-balance}
\end{align}
where $\Phi$ is the gravitational potential, $\mu_0$ is the vacuum permeability, and $\mathbf{B}$ is the magnetic field. In cylindrical polar coordinates $(\varpi, \phi, z)$, an axisymmetric magnetic field can generally be written
\begin{align}
  \mathbf{B} = \frac{1}{\varpi} \nabla \psi \times \hat{\boldsymbol{\phi}} + B_\phi \hat{\boldsymbol{\phi}} \:,
\end{align}
where $\psi = \psi(\varpi,z)$ is a poloidal flux function and $\hat{\boldsymbol{\phi}}$ is the unit vector in the azimuthal direction. If we additionally make the barotropic assumption $\nabla p \times \nabla \rho = 0$, we obtain (after some manipulation) the Grad-Shafranov equation
\begin{align}
  \Delta^* \psi + F \frac{\rmd F}{\rmd \psi} = -\mu_0 \varpi^2 \rho G \:, \label{eq:GSE} \\
  \shortintertext{where}
  \Delta^* \equiv \frac{\del^2}{\del z^2} - \frac{1}{\varpi} \frac{\del}{\del \varpi} + \frac{\del^2}{\del \varpi^2}
\end{align}
and $F = \varpi B_\phi$. It can be shown that $F = F(\psi)$ and $G = G(\psi)$, where $F$ and $G$ are as-yet unspecified functions.

The simplest choice that yields a non-trivial field solution is to set $F$ directly proportional to $\psi$ and $G$ to be a constant. This follows the approach of \citet{Prendergast1956}, who additionally had the insight to introduce the separation $\psi(r, \theta) = \Psi(r) \sin^2 \theta$, where $(r, \theta, \phi)$ are spherical polar coordinates. Setting $F = \lambda \psi$ and $G = -\beta/\mu_0$, where $\lambda$ and $\beta$ are constants, turns (\ref{eq:GSE}) into the linear, second-order, inhomogeneous ordinary differential equation (ODE)
\begin{align}
  \Psi'' - \left( \frac{2}{r^2} - \lambda^2 \right) \Psi = \beta \rho r^2 \:, \label{eq:GSE_linear}
\end{align}
which can be solved via a Green's function approach. Once the solution for $\Psi$ (which we will refer to as the radial flux function) is found, the magnetic field components can then be obtained as
\begin{align}
  \mathbf{B} &= (B_r, B_\theta, B_\phi) \nonumber \\
  &= \left( \frac{2}{r^2} \Psi(r) \cos \theta, -\frac{1}{r} \Psi'(r) \sin \theta, -\frac{\lambda}{r} \Psi(r) \sin \theta \right) \:.
\end{align}

\begin{figure}
  \centering
  \includegraphics[width=\columnwidth]{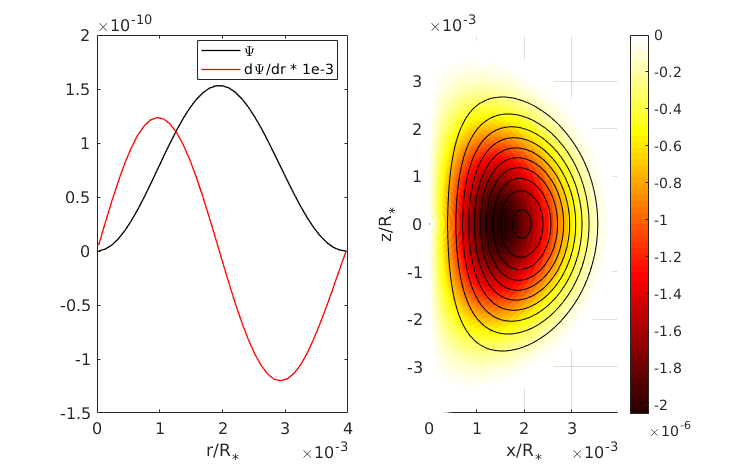}
  \includegraphics[width=\columnwidth]{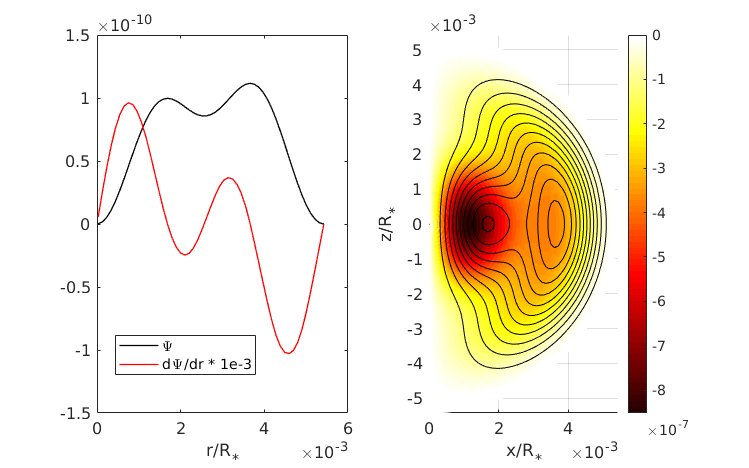}
  \includegraphics[width=\columnwidth]{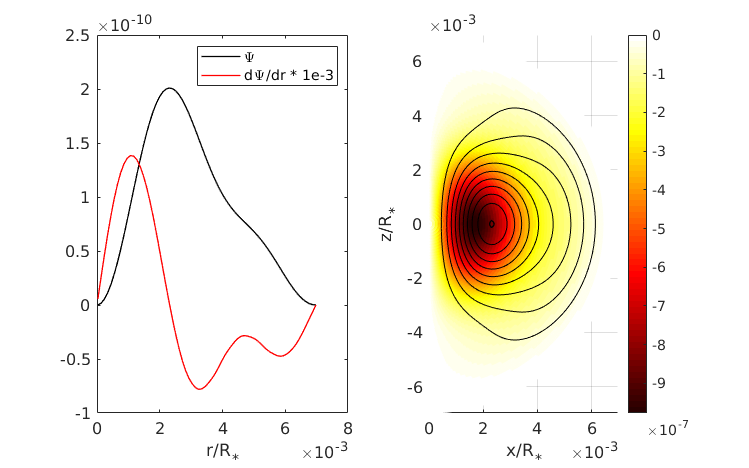}
  \caption{Prendergast solutions for Model B (young RGB), calculated for three field radii. From top to bottom, these are $R_f = 4 \times 10^{-3}$, $5.5 \times 10^{-3}$ and $7 \times 10^{-3}\,R_*$ (dashed lines in Figure \ref{fig:N_profiles_field_radii}, middle panel). The smallest $\lambda$ roots were chosen in each case, these being $1.71 \times 10^3$, $2.38 \times 10^3$ and $1.94 \times 10^3\,R_*^{-1}$, respectively. The panels on the left show the radial dependence of the radial flux function $\Psi(r)$ and its radial derivative (the latter scaled for clarity). On the right is a meridional half-section showing the poloidal projections of the field lines (black contours) and the strength of the toroidal component in colour. The field amplitude has been set such that $V_{A,\text{cen}} = V_{A,\text{crit}}$.}
  \label{fig:Prendergast_profile30}
\end{figure}

A non-trivial aspect of the solution to (\ref{eq:GSE_linear}) concerns the boundary conditions, of which there are three: $\Psi(0) = 0$, $\Psi(R_f) = 0$ and $\Psi'(R_f) = 0$, since all components of the field must vanish continuously to zero at the boundary to avoid infinite current sheets. Now (\ref{eq:GSE_linear}) is only a second-order ODE, and so a third degree of freedom must be invoked to satisfy the third boundary condition. This is provided through appropriate adjustment of $\lambda$. The general solution, given a background density profile $\rho(r)$ and field radius $R_f$, is
\begin{align}
  \Psi(r) &= \frac{\beta \lambda r}{j_1(\lambda R_f)} \left[ f_\lambda(r,R_f) \int_0^r \rho \xi^3 j_1(\lambda \xi) \rmd \xi \right. \nonumber \\
    &\quad \left. + j_1(\lambda r) \int_r^{R_f} \rho \xi^3 f_\lambda(\xi, R_f) \rmd \xi \right] \:, \label{eq:Prendergast_soln}
\end{align}
where $f_\lambda(r_1, r_2) \equiv j_1(\lambda r_2) y_1(\lambda r_1) - j_1(\lambda r_1) y_1(\lambda r_2)$ with $j_1$ and $y_1$ being spherical Bessel functions, and $\lambda$ satisfies
\begin{align}
  \int_0^{R_f} \rho \xi^3 j_1(\lambda \xi) \rmd \xi = 0 \:. \label{eq:lambda_cond}
\end{align}
Equations (\ref{eq:Prendergast_soln}) and (\ref{eq:lambda_cond}) define what we shall refer to as the \textit{Prendergast solution}. This is an ideal choice for modelling spherically confined equilibrium fields. It is dipole-like in appearance, but additionally contains a toroidal component of comparable strength to the poloidal one. Significant advantages it has over e.g.~a dipole field are that it is finite everywhere including at the origin, and vanishes smoothly at a finite radial distance $R_f$ in all components of $\mathbf{B}$. Figure \ref{fig:Prendergast_profile30} shows the Prendergast solutions obtained for the three $R_f$ values tested for Model B. Where multiple possible values of $\lambda$ existed satisfying (\ref{eq:lambda_cond}), the smallest value was chosen. The solutions match smoothly to the zero solution outside $R_f$.

We remark that the use of the barotropic assumption is not entirely self-consistent: while (\ref{eq:GSE_linear}) requires $\rho = \rho(r)$, the Lorentz force cannot be spherically symmetric which means that $p = p(r,\theta)$ to satisfy (\ref{eq:force-balance}). This leads to $\nabla p \times \nabla \rho \neq 0$. However, in the case of red giants we expect the departure of $p$ from sphericity to be very small, since thermal pressures greatly dominate over magnetic pressures $p_m = B^2/(2\mu_0)$ even for fields of critical strengths. The Lorentz force therefore has negligible effect on the equilibrium stellar structure, although it can still play a significant role in influencing gravity wave propagation. Departures from sphericity caused by magnetic deformation, which go as $\Delta R/R \sim p_m/p \sim 10^{-6}$, are thus very small. Hence $p \approx p(r)$, $\rho \approx \rho(r)$ to good approximation, and $\rho$ on the RHS of (\ref{eq:GSE_linear}) can be replaced by the non-magnetic (spherically symmetric) version. The dynamical consequences of non-sphericity are also minimal; for example, the associated time scale of meridional circulation $\tau_\text{circ} \sim \tau_\text{KH} p/p_m$ (where $\tau_\text{KH}$ is the Kelvin-Helmholtz time scale) is $\gtrsim 10^{11}\,$yr for critical field strengths for the models considered, greatly exceeding evolutionary time scales. We thus retain the barotropic approximation here.

\subsubsection{Choice of field strengths}\label{sec:vA}
The work of LP18 showed that the trapping phenomenon emerges when magnetic fields are strong enough to be dynamically significant, which occurs in parts of parameter space where Alfv\'{e}n wave frequencies and wavelengths become simultaneously comparable to those of gravity waves (i.e.~resonance occurs). That is, $\omega_A(\mathbf{k}) \sim \omega_g(\mathbf{k})$, where
\begin{align}
  \omega_A(\mathbf{k}) = \mathbf{k} \cdot \mathbf{V}_A \:,\quad \omega_g(\mathbf{k}) = \frac{k_\perp}{k} N \:, \label{eq:frequencies}
\end{align}
are the frequencies of Alfv\'{e}n and gravity waves, respectively. Here $k \equiv |\mathbf{k}|$, $\mathbf{V}_A$ is the Alfv\'{e}n velocity, and $k_\perp$ is the component of $\mathbf{k}$ perpendicular to the direction of stratification. For a mode of spherical harmonic degree $\ell$,
\begin{align}
  k_\perp = \frac{\sqrt{\ell(\ell+1)}}{r} \:,
\end{align}
and so for a wave frequency of $\omega$, resonant interactions would be expected to occur when the Alfv\'{e}n speed $V_A \equiv |\mathbf{V}_A| \sim V_{A,\text{crit}}$, where
\begin{align}
  V_{A,\text{crit}} \sim \frac{r}{N} \frac{\omega^2}{\sqrt{\ell(\ell+1)}} \:. \label{eq:vAcrit}
\end{align}
Up to a factor of two, this is the same expression obtained by \citet{Fuller2015}. In deriving this we have assumed alignment of $\mathbf{k}$ and $\mathbf{V}_A$, which means that $V_{A,\text{crit}}$ in the above expression is a lower bound for the onset of resonance. Note that $V_A \propto B$, so the field amplitude is equivalently controlled by the scaling of $V_A$. Near the very centre of a star it is the case that approximately $N \propto r$, with the proportionality factor $r/N$ observed to be $\sim 10^{-4}$, $5 \times 10^{-6}$ and $2 \times 10^{-7}$ (in units of $R_*/\omega_\text{dyn}$) for Models A, B and C, respectively. For typical values of $\omega$ and $\ell$ tested (see next section), we estimated critical Alfv\'{e}n speeds $V_{A,\text{crit}}$ to be $4 \times 10^{-3}$, $2 \times 10^{-4}$ and $8 \times 10^{-6}\,V_\text{dyn}$. See Table \ref{tab:profiles} for the dimensional versions of these values, and the corresponding critical field strengths computed as $B_\text{crit} = V_{A,\text{crit}} \sqrt{\mu_0 \rho_c}$, for each model. Associated central values of the plasma beta $p/p_m$ for each model are $3 \times 10^5$, $5 \times 10^8$ and $10^{12}$ for Models A, B and C respectively, and departures from sphericity $\Delta R/R \sim p_m/p$ are accordingly small. See that $V_{A,\text{crit}}$ is a quantity that depends entirely on the background properties of a star, and can be computed for a mode of given $\omega$ and $\ell$ without any recourse to ray tracing.

We tested a range of field strengths around the estimated critical value for each model. The field strength was controlled by setting the central Alfv\'{e}n speed $V_{A,\text{cen}}$ to chosen multiples of $V_{A,\text{crit}}$. For the middle of the three $R_f$ values of Models A and B, and the single $R_f$ value of Model C (see Section \ref{sec:Rf}), seven values were tested:
\begin{align}
  V_{A,\text{cen}} \in \{ 0.01, 0.03, 0.1, 0.3, 1, 3, 10 \} \times V_{A,\text{crit}} \:.
\end{align}
For the smaller and larger values of $R_f$ in Models A and B, three values were tested:
\begin{align}
  V_{A,\text{cen}} \in \{ 0.1, 1, 10 \} \times V_{A,\text{crit}} \:.
\end{align}
Note that although $V_{A,\text{cen}}$ is not exactly the same as the maximum $V_A$ (which tends to occur slightly away from the centre for the Prendergast solution), it is within several per cent of this.

\subsection{Wave parameters}\label{sec:om_ell}
Red giants are solar-like oscillators, meaning their oscillations are stochastically driven by convection occurring in the envelope. For solar-like oscillators, the frequency of maximum power $\nu_\text{max} = \omega_\text{max}/2\pi$ is observed to be strongly correlated with the dynamical frequency (large separation) \citep{Bedding2013}, and typically has a value near $\sim 10\,\nu_\text{dyn}$ for red giants \citep{Mosser2013}. With observations of sufficient quality, multiple modes can often be observed in the vicinity of $\nu_\text{max}$, spanning a broad envelope whose width may be several times $\nu_\text{dyn}$. We tested three values of $\omega$ for each of the models, these being
\begin{align}
  \omega \in \{ 8, 10, 12 \} \omega_\text{dyn} \:. \label{eq:omega}
\end{align}

For Models A and B, for each value of $\omega$, we additionally tested three values of the spherical harmonic degree $\ell$:
\begin{align}
  \ell \in \{ 1, 2, 3 \} \:. \label{eq:degree}
\end{align}
For Model C, only $\ell = 1$ was tested. Most asteroseismic observations cannot resolve the stellar disc; it is only the disc-integrated flux that is measured. Hence modes of high spherical harmonic degree, which have large numbers of angular nodes, suffer geometric cancellation and are difficult to detect. Only modes of low degree, typically those with $\ell \leq 3$, can be reliably identified in asteroseismic power spectra. Although $\ell = 0$ (radial modes) are also observable, gravity waves with non-trivial frequencies only exist for $\ell \geq 1$ since they necessarily involve some horizontal structure, and so we ignore $\ell = 0$ here.

%% METHODS
\section{Methods}\label{sec:methods}
\subsection{Magneto-gravity ray tracing}
\subsubsection{Hamilton's equations}
The trajectories of a Hamiltonian system evolve according to
\begin{align}
  \frac{\rmd \mathbf{x}}{\rmd t} = \nabla_\mathbf{k} H \:,\quad \frac{\rmd \mathbf{k}}{\rmd t} = -\nabla H \:, \label{eq:Hamilton}
\end{align}
where $t$ is time, $\mathbf{x}$ are the spatial coordinates, $\mathbf{k}$ are the conjugate momenta, and $H = H(\mathbf{x}, \mathbf{k}, t)$ is the Hamiltonian. When applied to problems of wave mechanics, the frequency $\omega = \omega(\mathbf{x}, \mathbf{k}, t)$ given by the dispersion relation of the mode in question takes the role of the Hamiltonian, and the trajectories correspond to the group-velocity paths traced by wave packets (rays) launched in the system. The position and wavevector of the ray are given by $\mathbf{x}$ and $\mathbf{k}$. This technique can be applied to problems where wavelengths are much smaller than background scales (i.e.~the Wentzel-Kramers-Brillouin-Jeffreys or WKBJ ansatz of fluid mechanics), a generally good approximation for solar-like oscillators, and is a useful method for visualising the paths of energy flow. Furthermore, it has an advantage over full-wave calculations in that it is computationally inexpensive to implement, regardless of the geometrical complexity of the problem, since it always just involves integrating a system of linear first-order ODEs, for which standard techniques exist.

Deep within evolved stellar cores, sound speeds greatly exceed those of other wave modes and so the role of acoustic effects on dynamics can be neglected. The relevant Hamiltonian is thus given by the magneto-gravity dispersion relation
\begin{align}
  \omega^2 = \omega_A^2 + \omega_g^2 = (\mathbf{k} \cdot \mathbf{V}_A)^2 + \kappa_\perp^2 N^2 \:, \label{eq:MG_DR}
\end{align}
which comes from applying a WKBJ treatment to the fluid equations. Here $\kappa_\perp \equiv k_\perp/k$.

The problem setup in this work is a three-dimensional sphere in which an axisymmetric field has been embedded. For a Prendergast field the radial and angular dependencies separate, while the remaining background quantities are assumed to be spherically symmetric, so it is most convenient to recast Hamilton's equations in spherical polar coordinates. In this case we seek to evolve the six quantities $r$, $\theta$, $\phi$, $k_r$, $k_\theta$ and $k_\phi$. Substituting the dispersion relation (\ref{eq:MG_DR}) for the Hamiltonian in (\ref{eq:Hamilton}) leads to (see Appendix \ref{sec:rt_derivation})
\begin{align}
  \frac{\rmd r}{\rmd t} &= \frac{\omega_A}{\omega} V_{Ar} - \frac{N^2 \kappa_\perp^2 \kappa_r}{\omega k} \:, \label{eq:drdt} \\
  \frac{\rmd \theta}{\rmd t} &= \frac{\omega_A}{\omega} \frac{V_{A\theta}}{r} + \frac{N^2 \kappa_\theta \kappa_r^2}{\omega k r} \:, \label{eq:dthetadt} \\
  \frac{\rmd \phi}{\rmd t} &= \frac{\omega_A}{\omega} \frac{V_{A\phi}}{r \sin \theta} + \frac{N^2 \kappa_\phi \kappa_r^2}{\omega k r \sin \theta} \:, \label{eq:dphidt}
\end{align}
\begin{align}
  \frac{\rmd k_r}{\rmd t} &= \frac{\omega_A}{\omega} \left( \frac{k_\theta V_{A\theta} + k_\phi V_{A\phi}}{r} - \mathbf{k} \cdot \frac{\del \mathbf{V}_A}{\del r} \right) + \frac{N \kappa_\perp^2}{\omega} \left( \frac{N \kappa_r^2}{r} - \frac{\rmd N}{\rmd r} \right) \:, \label{eq:dkrdt} \\
  \frac{\rmd k_\theta}{\rmd t} &= -\frac{1}{r} \frac{\omega_A}{\omega} \left( k_\theta V_{Ar} - k_\phi V_{A\phi} \cot \theta + \mathbf{k} \cdot \frac{\del \mathbf{V}_A}{\del \theta} \right) \nonumber \\
  &\qquad + \frac{N^2 \kappa_r}{r \omega} \left( \kappa_\theta \kappa_\perp^2 + \kappa_\phi^2 \kappa_r \cot \theta \right) \:, \label{eq:dkthetadt} \\
  \frac{\rmd k_\phi}{\rmd t} &= -\frac{k_\phi}{r} \frac{\omega_A}{\omega} \left( V_{Ar} + V_{A\theta} \cot \theta \right) + \frac{N^2 \kappa_\phi \kappa_r}{r \omega} \left( \kappa_\perp^2 - \kappa_\theta \kappa_r \cot \theta \right) \:, \label{eq:dkphidt}
\end{align}
where $\mathbf{k} = (k_r, k_\theta, k_\phi)$, $\boldsymbol{\kappa} = \mathbf{k}/k = (\kappa_r, \kappa_\theta, \kappa_\phi)$, $\kappa_\perp^2 = \kappa_\theta^2 + \kappa_\phi^2$, and $\mathbf{V}_A = (V_{Ar}, V_{A\theta}, V_{A\phi})$. Equations (\ref{eq:drdt})--(\ref{eq:dkphidt}) are the magneto-gravity ray tracing equations used in this work. Note that they assume $\del/\del \phi \equiv 0$ and $N = N(r)$.

\subsubsection{Integration}
Equations (\ref{eq:drdt})--(\ref{eq:dkphidt}) were integrated forward in time using a fourth-order Runge-Kutta scheme to obtain the ray trajectories. Note that the scheme does not automatically conserve $\omega$, which should be a constant of motion since the Hamiltonian has no explicit time dependence. Inspection of $\omega$ therefore serves as an independent check of accuracy. Preliminary tests of the integration routine on smooth, infinitely differentiable backgrounds showed conservation of $\omega$ to within one part in $10^{16}$, i.e.~to machine precision. However, conservation of $\omega$ was more problematic when the MESA profiles were used; this is likely a consequence of non-smoothness, since the values were interpolated from a finite grid. However, it is crucial that the wave frequency stays unchanged, since it is a constant of motion in the physical problem and influences important aspects of the propagation, such as the locations of turning points. Details of measures to enforce this can be found in Appendix \ref{sec:omega_conservation}.

A time step size of 0.001 was used (time is measured in units of $\omega_\text{dyn}^{-1}$), which was such that a five-fold decrease in this led to no noticeable improvement in $\omega$ conservation. Rays were integrated up to a specified maximum duration, this being 100, 400 and 2000 time units for Models A, B and C respectively. The difference in these set durations was due to the different group velocities and thereby crossing times of the rays; empirically, it was observed that rays in Models A, B and C took roughly 20, 80 and 400 time units each to complete one crossing of the g-mode cavity. The maximum durations were set so as to allow for 4--5 bounces of a ray between inner and outer turning points, to give enough time to make a reliable classification into `trapped' or `reflected' (see Section \ref{sec:classification}).

\subsubsection{Ray initialisation}
For each model and combination of $V_{A,\text{cen}}$, $R_f$, $\omega$ and $\ell$ (see Sections \ref{sec:Rf}, \ref{sec:vA} and \ref{sec:om_ell}), 1200 rays were launched, all from the same radial distance $r_0$. It is to be noted that the actual value of the launch radius $r_0$ does not matter, owing to the way the initial wavevector $\mathbf{k}_0$ is constructed (more detail below). In practice, the launch radii were $r_0 = 0.04\,R_*$, $0.01\,R_*$ and $2.1 \times 10^{-3}\,R_*$ for Models A, B and C, respectively (red lines in Figure \ref{fig:N_profiles_field_radii}), which lie outside the field regions. The 1200 rays were launched from 30 colatitudes $\theta_0$, spread evenly between the poles, and for each $\theta_0$ they were launched with 40 different ``polarisations'' (orientations of $\boldsymbol{\kappa}_\perp$), spread evenly round a circle. This was parametrised through the position angle $\alpha$, defined to be $\alpha = \tan^{-1}(k_\theta/k_\phi)$, i.e.~clockwise angle from the $\hat{\boldsymbol{\phi}}$ direction. This was intended to mimic the arbitrariness of propagation direction for waves excited in a spherically symmetric envelope; subsequent averaging over all directions then yields an effective quantity that might be associated to a global mode.

Initialisation of a ray requires starting values for all six quantities: $r_0$, $\theta_0$, $\phi_0$, $k_{r0}$, $k_{\theta0}$ and $k_{\phi0}$. The values of $r_0$ and $\theta_0$ are independently chosen as above, and $\omega$ and $\ell$ are also independently chosen as described in Section \ref{sec:om_ell}. We then determine
\begin{align}
  k_{r0} &= \sqrt{k_0^2 - k_{\perp0}^2} \:, \\
  k_{\theta0} &= k_{\perp0} \sin \alpha \:, \\
  k_{\phi0} &= k_{\perp0} \cos \alpha \:,
\end{align}
where $k_{\perp0} = \sqrt{\ell(\ell+1)}/r_0$ and $k_0 = k_{\perp0}N(r_0)/\omega$. Note that since the setup is axisymmetric, the value of $\phi_0$ is inconsequential; for simplicity, we set $\phi_0 = \alpha$.

To explain further as to why the choice of launch radius does not matter, let us consider propagation of a gravity wave in a region of zero field. This is governed by the dispersion relation $\omega = k_\perp N / k$, where $k_\perp = \sqrt{\ell(\ell+1)}/r$. For chosen values of $\omega$ and $\ell$, if a ray were to be launched from $r = r_1$ then its initial $k_r$ value would be given by $k_{r1}^2 = [N(r_1)^2/\omega^2 - 1] [\ell(\ell+1)]/r_1^2$. Suppose this ray is allowed to propagate for some time until it reaches a different radius, $r = r_2$. Because the Hamiltonian possesses time independence and spherical symmetry, $\omega$ and $\ell$ are constants of motion, and hence the value of $k_r$ at the later time would be $k_{r2}^2 = [N(r_2)^2/\omega^2 - 1] [\ell(\ell+1)]/r_2^2$. By construction, this is exactly the value of $k_r$ that would have been assigned to the ray, had it been launched from $r = r_2$ instead. Identical arguments apply to $k_\perp$. Thus pure gravity rays have no memory of what radius they were launched from, because for fixed $\omega$ and $\ell$ they will always have the same $k_r$ and $k_\perp$ at given $r$. However, this property fails once the rays enter the field region. Thus as long as the launch location is outside the field region, $r_0$ can be chosen arbitrarily with no impact on the results.

\subsection{Analysis}
\subsubsection{Quality control}\label{sec:quality}
As detailed in Appendix \ref{sec:omega_conservation}, we attempted to enforce conservation of $\omega$ through an iterative root-finding procedure at each time step. However, this did not always converge. Rays for which convergence failed could be identified through sudden excursions in $\omega$; where they occurred, these were often by several per cent or larger. We thus chose to discard all rays whose $\omega$ values fluctuated by more than 1\% from the initial value. For Model A, this accounted for less than 1\% of all rays, while for Model B this accounted for 5\%. In the case of Model C, whose background profiles were least smooth, 28\% of rays were discarded under this criterion.

\subsubsection{Ray classification}\label{sec:classification}
The remaining rays were then classified as being either `reflected' or `trapped', as follows. If at any point a ray crossed $r = R_f$ going out, it was considered to be `reflected'. Otherwise, if the final wavenumber exceeded the initial wavenumber $k_0$ by a factor of 100, it was classified as `trapped'. A small number of rays were left unclassified; this accounted for less than 0.1\% of rays for Model A, and was slightly higher for the others, being 3\% for Model B and 5\% for Model C.

The reflected and trapped fractions $f_R$ and $f_T$ were calculated from the two sets of rays, $\mathcal{R}$ (reflected) and $\mathcal{T}$ (trapped), by weighting each ray by its areal contribution $w_i = \sin \theta_0[i]$, where $\theta_0[i]$ is the launch colatitude of ray $i$, as
\begin{align}
  f_R = \frac{1}{W_C} \sum_{i \in \mathcal{R}} w_i \:,\quad f_T = \frac{1}{W_C} \sum_{i \in \mathcal{T}} w_i \:,
\end{align}
where $W_C = \sum_{i \in \mathcal{R} \cup \mathcal{T}} w_i$ is the sum of weights of successfully classified rays. By weighting in this manner, the result is effectively a geometrical average over the surface of the star, and relates to the trapping fraction experienced by a global mode of given $\omega$ and $\ell$.

\subsubsection{Uncertainty estimation}\label{sec:uncertainties}
Owing to the very large number of rays launched and analysed (hundreds of thousands in total), it was not practical to follow up individually on the unclassified and non-convergent rays. Inspection of a number of unclassified rays showed that many of these appeared to be on large-scale delocalised orbits, with the wavenumber not diverging and yet the ray not having left the magnetised region. It is anticipated that many of these would have been classifiable given more integration time. In the case of non-convergent rays, the failure at certain points in time to converge to the appropriate root of (\ref{eq:kr_quartic}) meant that those rays were observed to ``reset'' their trajectory (abruptly change in $k_r$), making subsequent classification unreliable.

These rays represented a source of uncertainty in the calculations of $f_R$ and $f_T$. Our approach to incorporating and expressing this was to compute the absolute upper and lower bounds of $f_R$ (equivalently, the lower and upper bounds of $f_T = 1 - f_R$) by assuming that all unclassified and non-convergent rays fell entirely within one or the other class. That is, letting $\mathcal{U}$ be the combined set of unclassified/non-convergent rays, we computed
\begin{align}
  f_R^\text{max} = \frac{1}{W_\text{tot}} \sum_{i \in \mathcal{R} \cup \mathcal{U}} w_i \:,\quad f_T^\text{max} = \frac{1}{W_\text{tot}} \sum_{i \in \mathcal{T} \cup \mathcal{U}} w_i \:,
\end{align}
where $W_\text{tot} = \sum_{i \in \mathcal{R} \cup \mathcal{T} \cup \mathcal{U}} w_i$, and accordingly
\begin{align}
  f_R^\text{min} = 1 - f_T^\text{max} \:,\quad f_T^\text{min} = 1 - f_R^\text{max} \:.
\end{align}
The error bars shown in all plots of $f_R$ and $f_T$ are the intervals $[f_R^\text{min}, f_R^\text{max}]$ and $[f_T^\text{min}, f_T^\text{max}]$.

%% RESULTS
\section{Results}\label{sec:results}
\subsection{Trapping \& reflection}
\begin{figure}
  \centering
  \includegraphics[clip=true, trim=0cm 0.5cm 0cm 0.5cm, width=0.9\columnwidth]{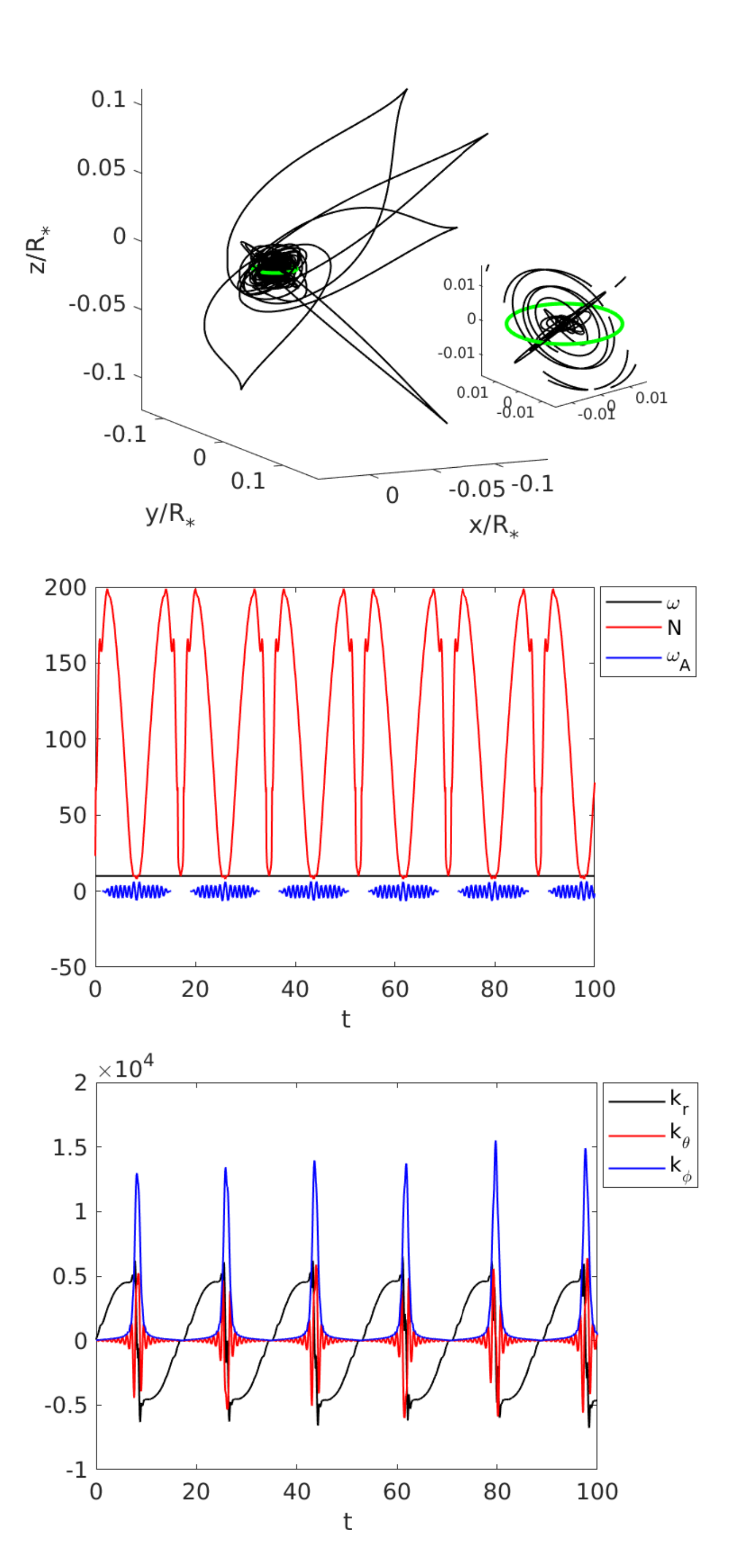}
  \caption{An example of a reflected ray, showing the trajectory plotted in three dimensions (top), the wave, buoyancy and Alfv\'{e}n frequencies as a function of time (middle), and the wavevector components as a function of time (bottom). This ray was computed for Model A (subgiant), with the field imposed within $R_f = 0.016\,R_*$ (green circle). The field strength was set such that $V_{A,\text{cen}} = 1.3 \times 10^{-3}\,V_\text{dyn}$ ($0.3\,V_{A,\text{crit}}$). The ray was launched from $(r_0, \theta_0) = (0.04\,R_*, 99^\circ)$ with $\omega = 10\,\omega_\text{dyn}$, $\ell = 1$ and $\alpha = \tan^{-1}(k_\theta/k_\phi) = 40.5^\circ$. The inset plot in the top panel shows a zoom-in to the field region, where for clarity only the portion of the trajectory up to $t = 20$ has been plotted.} For a 3D representation of the ray trajectory, see rotating animation in the online supplementary material.
  \label{fig:ray137}
\end{figure}

\begin{figure}
  \centering
  \includegraphics[clip=true, trim=0cm 0.5cm 0cm 0.5cm, width=0.9\columnwidth]{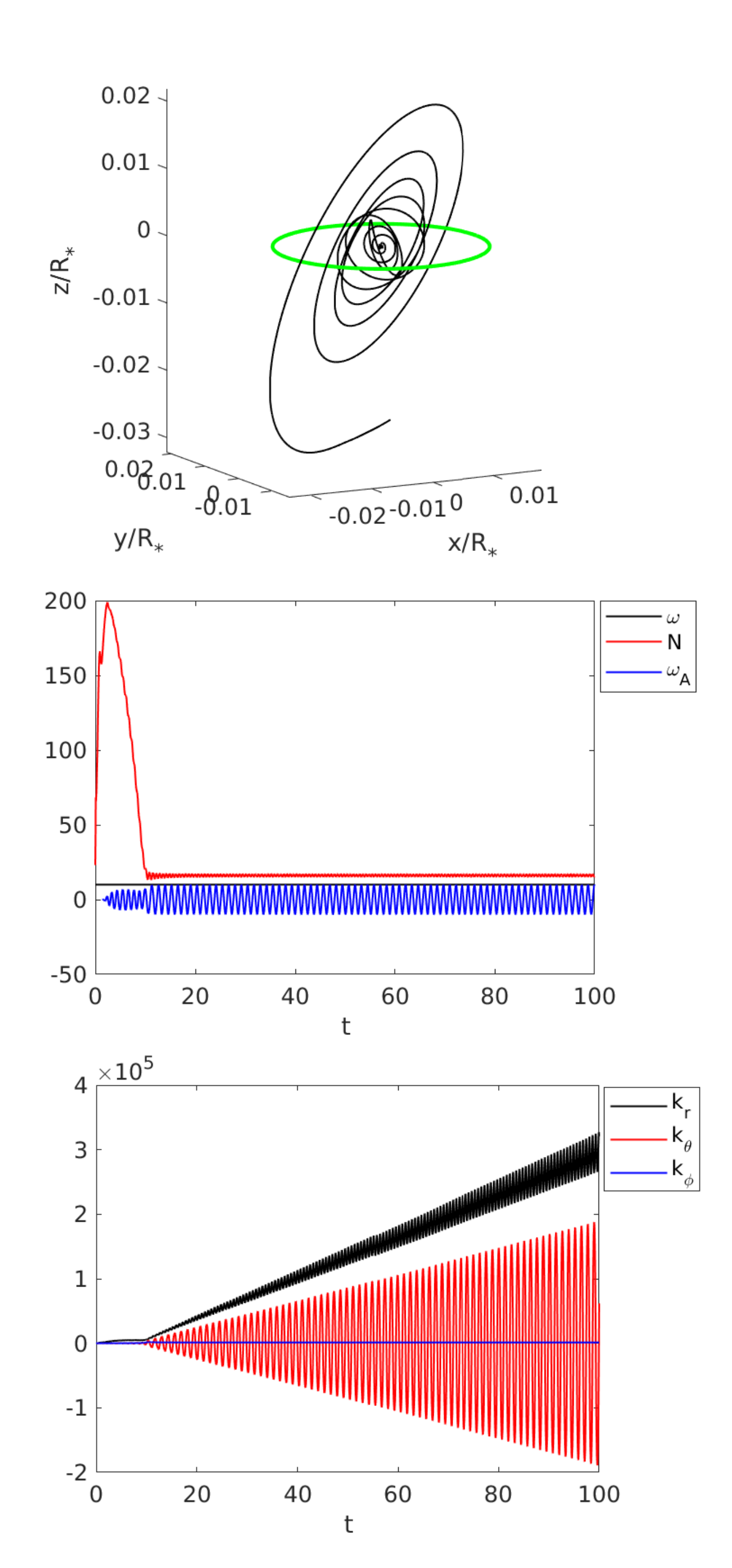}
  \caption{As for Figure \ref{fig:ray137}, but with a launch colatitude and position angle of $\theta_0 = 141^\circ$ and $\alpha = 58.5^\circ$, respectively. The remaining parameters ($V_{A,\text{cen}}$, $R_f$, $\omega$, $\ell$ and $r_0$) are identical to those of Figure \ref{fig:ray137}. Unlike the previous ray, this ray is instead trapped, exhibiting a wavenumber that diverges with time. For a 3D representation of the ray trajectory, see rotating animation in the online supplementary material.}
  \label{fig:ray204}
\end{figure}

The work of LP18 showed trapping and reflection occurring for magneto-gravity rays in a periodic Cartesian box with a helical magnetic field. Figures \ref{fig:ray137} and \ref{fig:ray204} show those same phenomena occur in a more realistic geometry. In Figure \ref{fig:ray137} (top) the trajectory of a single ray computed for Model A is shown. This undergoes multiple passages into and out of the magnetised region, whose boundary is marked with a green circle and whose amplitude is such that $V_{A,\text{cen}} = 0.3 V_{A,\text{crit}}$. Spherical symmetry is noticeably broken by the field, which can be inferred through the significantly out-of-plane motion of the ray (this can be more easily seen in the movie available in the online supplementary material). The middle panel of Figure \ref{fig:ray137} plots $\omega$ (black), $N$ (red) and $\omega_A$ (blue) as a function of time. Note that when the ray is outside the field region, $\omega_A$ is not plotted, but its value there is simply zero. The bottom panel shows the evolution of the wavevector components over time. While some periodic exchanges are observed to occur between the various components, particularly near the turning points, it can be seen that the overall wavenumber remains stable and bounded with time.

The situation is very different for the ray shown in Figure \ref{fig:ray204}. This trajectory was also computed for Model A, with the field strength, field radius, launch radius, value of $\omega$ and $\ell$ being identical to that in Figure \ref{fig:ray137}. However, the ray in Figure \ref{fig:ray204} is trapped, exhibiting an inspiralling path that asymptotes towards and eventually localises about a certain flux surface. Crucially, the only differences between the two rays were the launch colatitude $\theta_0$ and initial polarisation position angle $\alpha$. From the plot of frequencies versus time (Figure \ref{fig:ray204}, middle), it can be seen that from about $t = 10$ the ray enters a regime where $\omega_A$ dominates the contribution to $\omega$. Here the ray is localised to a small range of radial distances (a shell) near the centre, but short of the inner turning point of the g-mode cavity. The bottom panel shows a clear divergence of the wavenumber with time. This behaviour for trapped rays has been previously noted by LP18, and independently by \citet{Valade2018}, and is thought to reflect the process of phase mixing (see LP18 for more discussion). The irreversible shrinkage of spatial scales means that in reality such waves would eventually be dissipated.

\begin{figure}
  \centering
  \includegraphics[clip=true, trim=0cm 0.5cm 0cm 0.5cm, width=0.9\columnwidth]{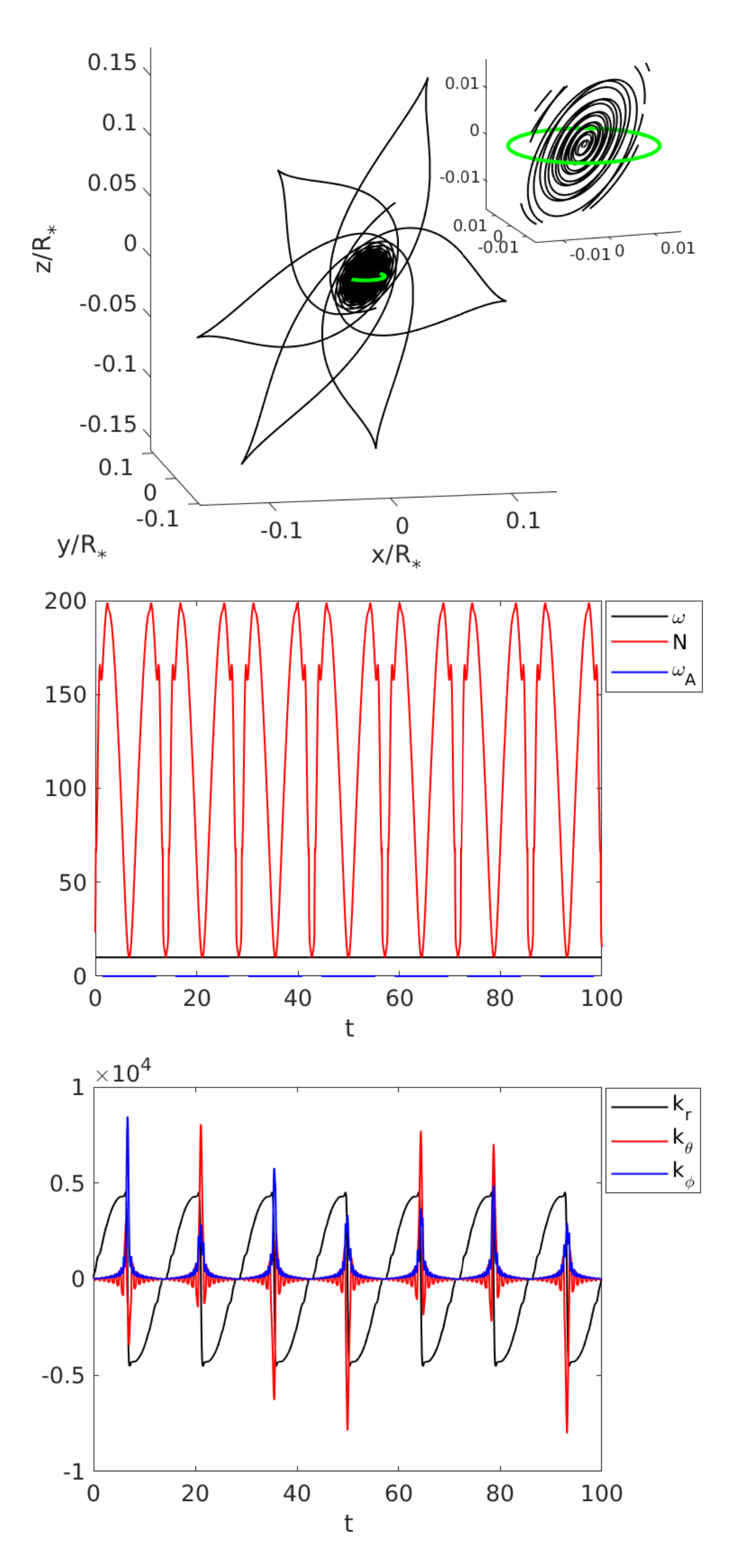}
  \caption{As for Figure \ref{fig:ray137}, showing the trajectory of the ray with identical launch parameters as that in Figure \ref{fig:ray204}, with the sole difference that now $V_{A,\text{cen}} = 0$. The green circle has the same radius as that in Figures \ref{fig:ray137} and \ref{fig:ray204}, but otherwise the field has been switched off. Unlike the other two rays, the trajectory of this ray lies in a single plane (a consequence of spherical symmetry). For a 3D representation of the ray trajectory, see rotating animation in the online supplementary material.}
  \label{fig:ray204_nofield}
\end{figure}

For comparison, in Figure \ref{fig:ray204_nofield} we show the trajectory of a third ray, launched with identical parameters as the trapped one in Figure \ref{fig:ray204}, but with the field turned off. In this case, no trapping occurs, and the ray behaves as one expects a pure gravity wave to: it bounces repeatedly between inner and outer turning points of the g-mode cavity (where $\omega = N$), and the orbit lies in a single plane. The middle and bottom panels can be seen to bear close qualitative resemblance to those of Figure \ref{fig:ray137}, with the exception that now $\omega_A$ is zero everywhere. Inspection of the quantity $r k_\perp$, which should be conserved under conditions of spherical symmetry (i.e.~zero field), shows that this is constant to within one part in $10^8$. This is also true for the unmagnetised portions of the trajectory in Figure \ref{fig:ray137}. In addition, we verified that the quantity $r k_\phi \sin \theta$ (the $z$-component of the angular momentum, which should be conserved under conditions of axisymmetry) is constant to within one part in $10^8$ for all three cases. Note that the fine-scale oscillatory structure in $k_\theta$ arises from the rotation of the wavevector as the ray spirals around the centre, which produces a quasi-sinusoidal variation of its projection in the $\hat{\boldsymbol{\theta}}$-direction.

To examine the possible long-term evolution of the nature of the orbits, we integrated a small subset of rays for a 10-fold longer duration, sufficient to allow for $50+$ reflections. Interestingly, it appears that there is no crossover between the two groups of rays, i.e.~if a ray is reflected once then it is always reflected upon return, and furthermore there is no systematic evolution in the periodic bounce pattern even on the longer time scale. It can already be seen from Figure \ref{fig:ray137} that in this case the periodicity persists for at least six bounces, and extending the total integration time to 1000 time units shows a total of 55 similar bounces occurring; the periodic pattern appears to be unceasing. For rays that are trapped, they remain so, and their wavenumber divergence never reverses. The explanation behind this is not clear, since one might na\"{i}vely expect it to be possible for a ray to undergo several reflections and then subsequently become trapped. However, there appear to be no such examples of this. A useful implication is that if reflected rays were ever to become trapped (over perhaps much longer timescales), then this is likely to occur only after tens to hundreds of bounces, meaning that their contribution to the energy loss rate would be 1--2 orders of magnitude smaller than those of the trapped group, and can be neglected in comparison.

\subsection{Dependence on field strength}
\begin{figure}
  \centering
  \includegraphics[clip=true, trim=0cm 1cm 0cm 0cm, width=0.95\columnwidth]{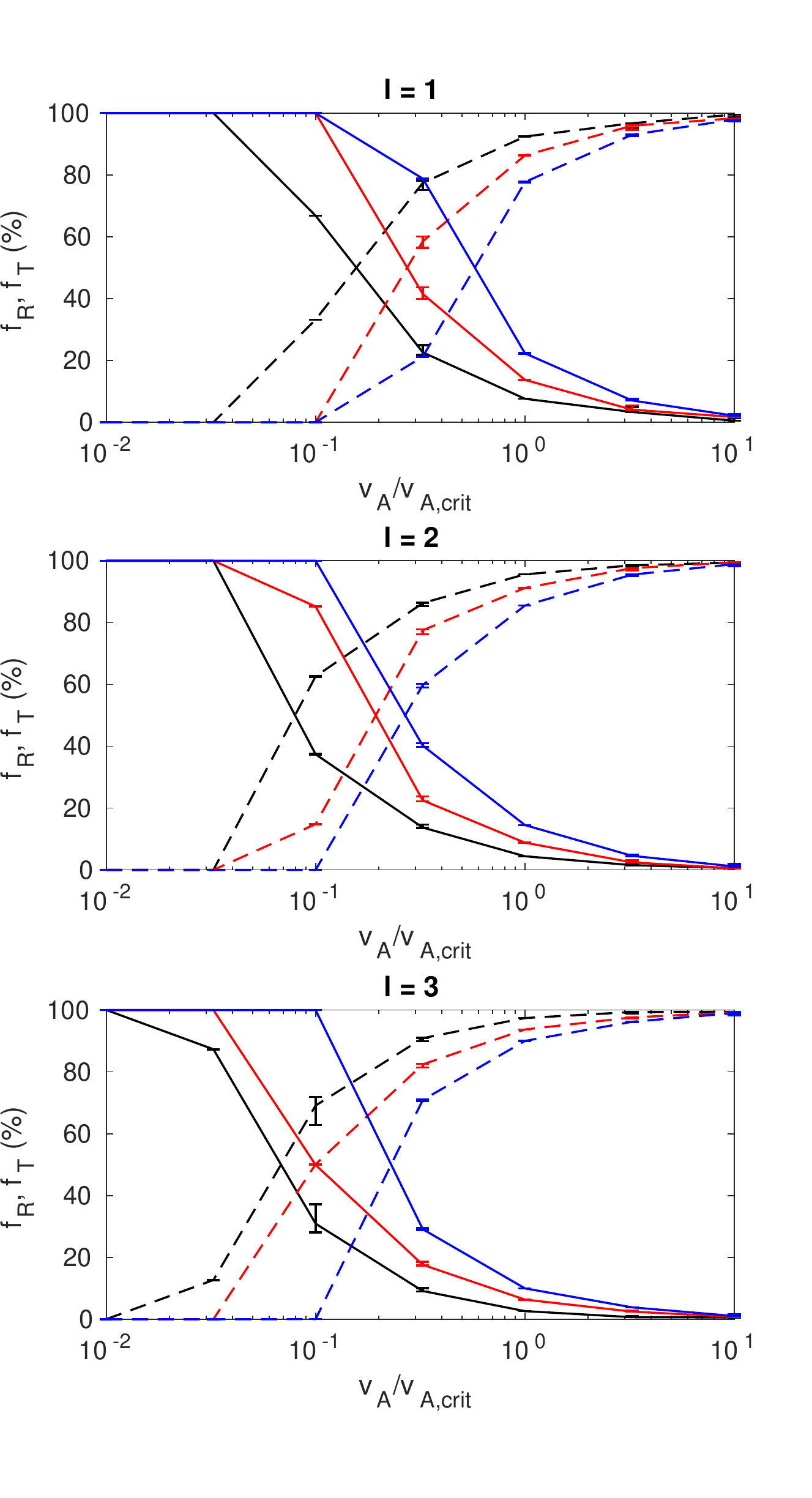}
  \caption{Fraction of rays (weighted by their areal contribution) that were reflected or trapped, for Model A (subgiant), as a function of field strength expressed as a fraction of the critical strength. Reflected fractions $f_R$ are shown with solid lines, while trapped fractions $f_T$ (100\% minus reflected) are shown with dashed lines. Black, red and blue correspond respectively to $\omega = 8$, 10 and 12\,$\omega_\text{dyn}$, and the different panels correspond to different spherical degrees $\ell$. Error bars indicate absolute upper and lower bounds as discussed in Section \ref{sec:uncertainties}.}
  \label{fig:profile26_mgres_om_l}
\end{figure}

\begin{figure}
  \centering
  \includegraphics[clip=true, trim=0cm 1cm 0cm 0cm, width=0.95\columnwidth]{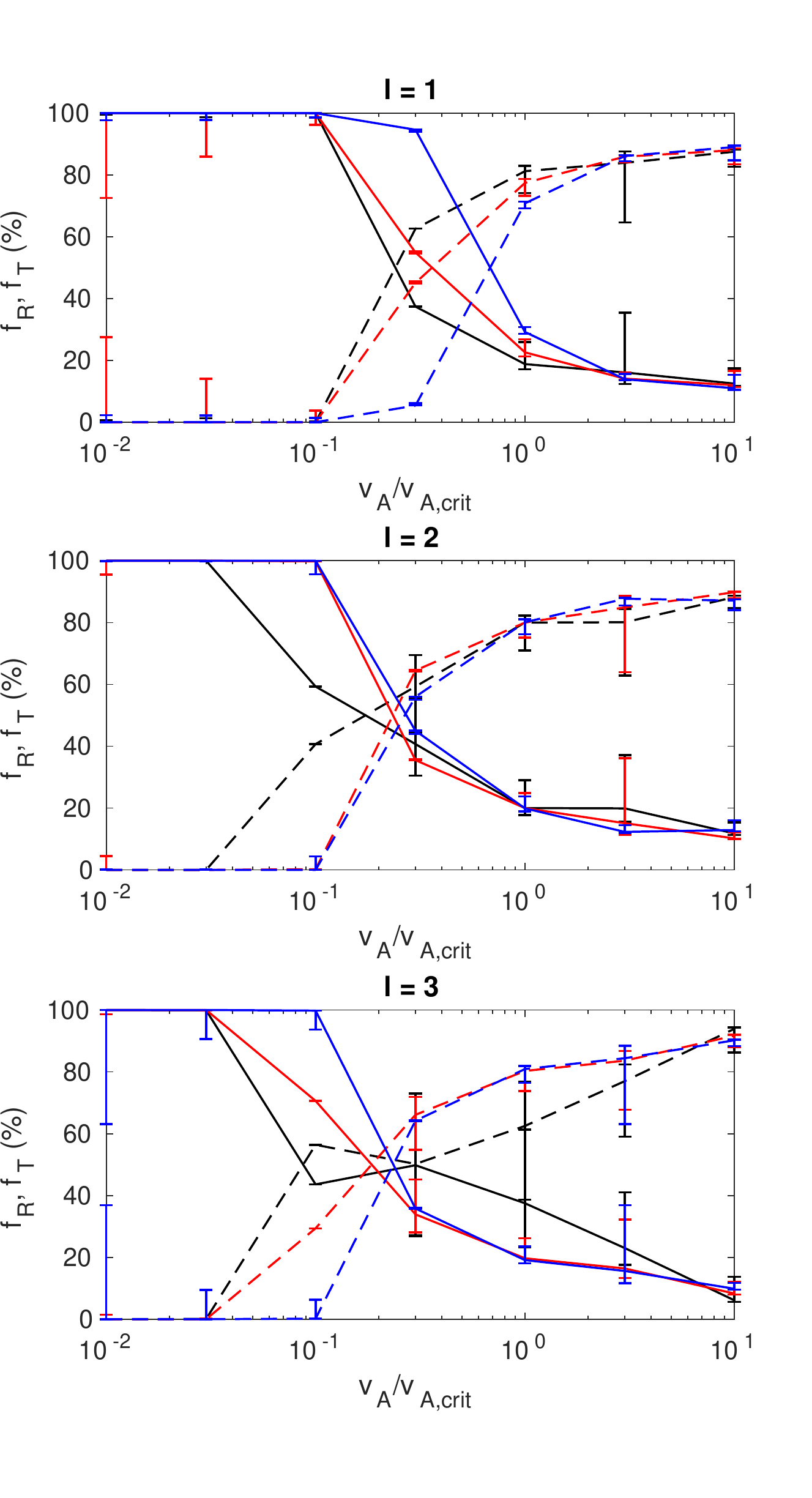}
  \caption{As for Figure \ref{fig:profile26_mgres_om_l}, but for Model B (young RGB).}
  \label{fig:profile30_mgres_om_l}
\end{figure}

Figures \ref{fig:profile26_mgres_om_l} and \ref{fig:profile30_mgres_om_l} show the trapped and reflected fractions as a function of field strength, for Models A and B, respectively. It can be seen that the transition from 100\% reflection (which one would get in the absence of a field) to maximal trapping occurs gradually over several orders of magnitude in the field strength. It also appears as though only the youngest model (A) approaches 100\% trapping, at least for the range of field strengths tested. For Model B, which is more evolved, the trapping fraction appears to saturate around 80--90\%. In each case, the crossover point occurs when $V_{A,\text{cen}}$ approaches $V_{A,\text{crit}}$ within a factor of several.

\subsection{Dependence on spherical degree}
The results for different values of $\ell$ are shown as separate panels in Figures \ref{fig:profile26_mgres_om_l} and \ref{fig:profile30_mgres_om_l}. It can be seen that the crossover point shifts to lower field strengths as $\ell$ increases, meaning that smaller field strengths would be required to produce the same rate of trapping for higher $\ell$. This is unsurprising, since from (\ref{eq:vAcrit}), $V_{A,\text{crit}} \propto 1/\sqrt{\ell(\ell+1)}$. It also means that for a given field strength and frequency, larger values of $\ell$ are expected to experience higher rates of trapping and consequent dissipation.

\subsection{Dependence on wave frequency}\label{sec:dep_om}
The different values of $\omega$ are shown as the different colours in Figures \ref{fig:profile26_mgres_om_l} and \ref{fig:profile30_mgres_om_l}. The crossover point occurs at higher field strengths for larger $\omega$, which can also be understood from (\ref{eq:vAcrit}) since $V_{A,\text{crit}} \propto \omega^2$. This implies that for a given field strength and spherical harmonic degree, if the field strength is near the crossover point, one might expect to observe a gradation in the rate of trapping (and thereby dissipation) with $\omega$, this being larger for smaller $\omega$. Interestingly, several stars of this description have been reported in the literature, including KIC 8561221 (``Droopy''), KIC 7746983, and KIC 6975038 \citep{Garcia2014, Mosser2017a}.

\subsection{Dependence on size of field region}
\begin{figure}
  \centering
  \includegraphics[clip=true, trim=0cm 1cm 0cm 0cm, width=0.95\columnwidth]{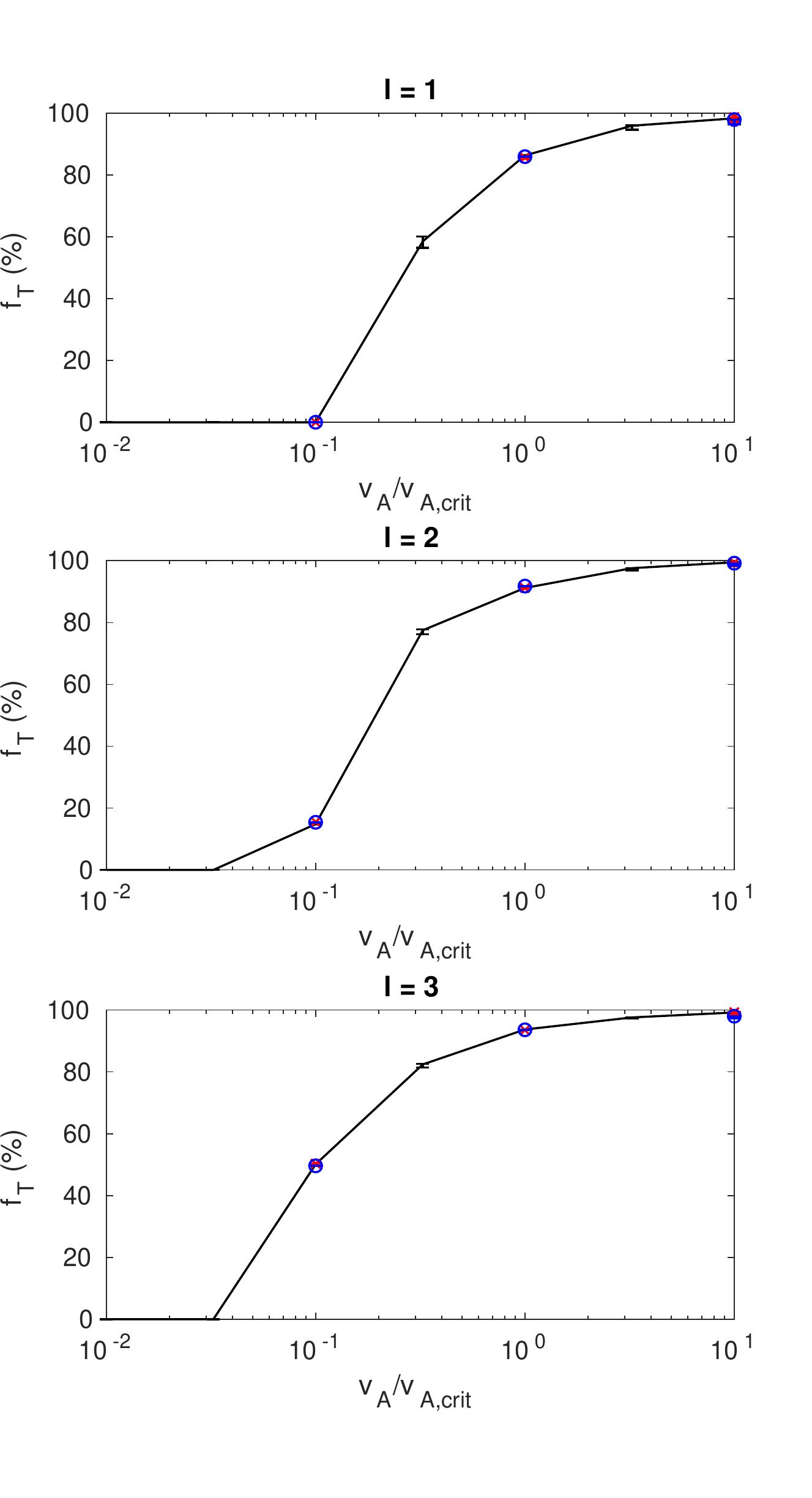}
  \caption{Trapped fraction $f_T$ as a function of field strength, for Model A (subgiant), comparing different field radii (Figure \ref{fig:N_profiles_field_radii}, top). The intermediate value ($R_f = 0.016\,R_*$) is shown with a black solid line, while the overlaid red crosses and blue circles correspond to $R_f = 0.012$ and $0.2\,R_*$. The wave frequency here is $\omega = 10\,\omega_\text{dyn}$. Note that for the smaller and larger values of $R_f$, only field strengths 0.1, 1, and 10 times the critical value were tested.}
  \label{fig:profile26_mgres_Rf_l}
\end{figure}

\begin{figure}
  \centering
  \includegraphics[clip=true, trim=0cm 1cm 0cm 0cm, width=0.95\columnwidth]{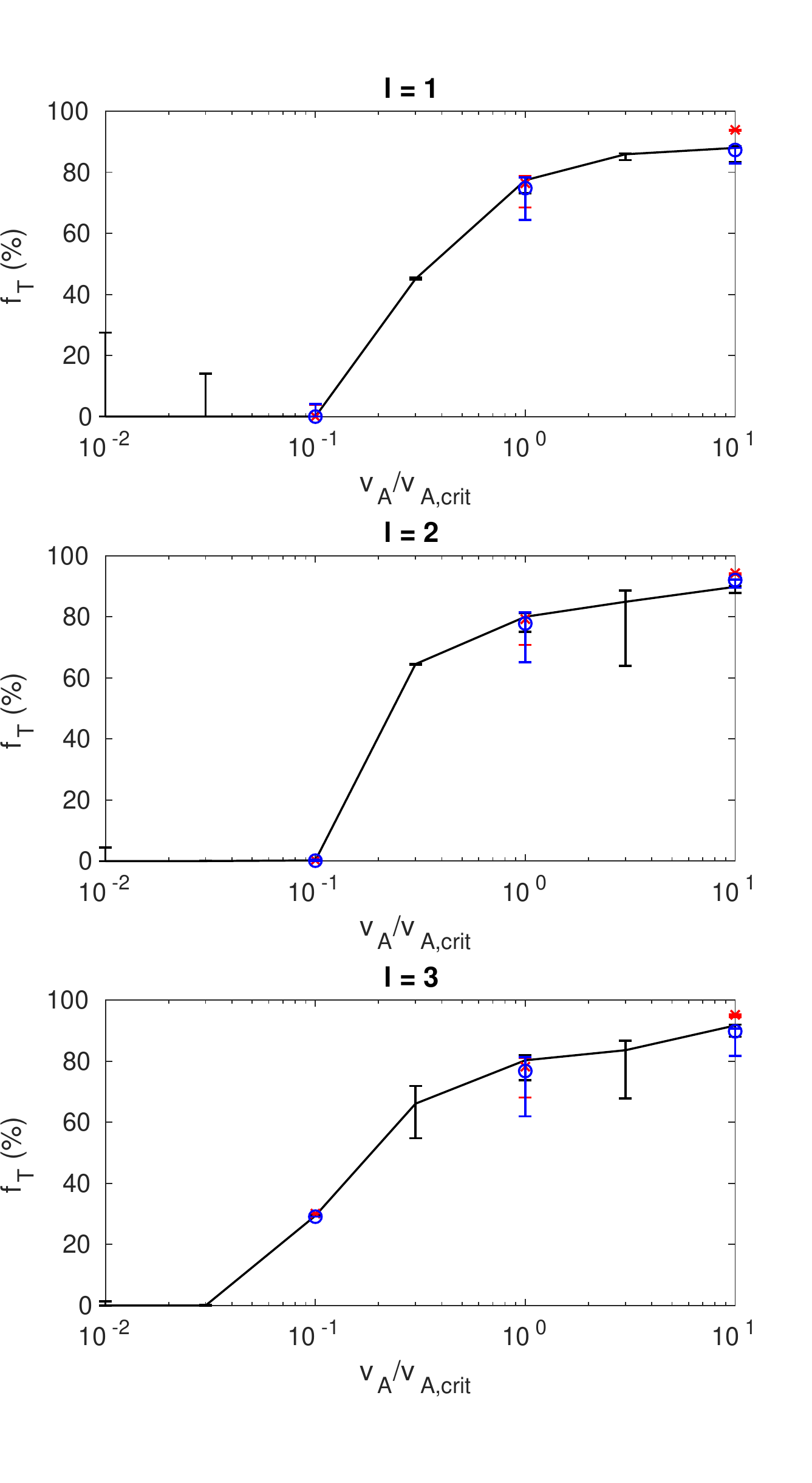}
  \caption{As for Figure \ref{fig:profile26_mgres_Rf_l}, but for Model B (young RGB). In this case the three values of $R_f$ are $5.5 \times 10^{-3}\,R_*$ (black solid line), $4 \times 10^{-3}\,R_*$ (red crosses) and $7 \times 10^{-3}\,R_*$ (blue circles). These are marked in Figure \ref{fig:N_profiles_field_radii} (middle) and the magnetic field configurations are shown in Figure \ref{fig:Prendergast_profile30}.}
  \label{fig:profile30_mgres_Rf_l}
\end{figure}

Figures \ref{fig:profile26_mgres_Rf_l} and \ref{fig:profile30_mgres_Rf_l} plot the trapped fractions versus field strength for Models A and B. The black lines join the seven points for the middle value of $R_f$ tested in each model. Additionally, three points in red and three points in blue are included showing the corresponding values for the smaller and larger values of $R_f$. For Model A (Figure \ref{fig:profile26_mgres_Rf_l}) in particular, the different points and their error bars lie nearly on top of one another. This implies that the radius of the field region has very little impact on the trapped/reflected fractions. This is despite nearly a factor of two variation in $R_f$ (about a factor of 5 variation in magnetised volume). It suggests that the effective solid angle subtended by trapping regions is more or less preserved as the radius of the field region is scaled up or down.

Note that the internal geometries of the twisted tori differ slightly between Prendergast solutions for different $R_f$. This can be seen in Figure \ref{fig:Prendergast_profile30}, where $\Psi$ and $\Psi'$ (which relate to different components of $\mathbf{B}$) have different numbers/locations of maxima/minima and radial crossings. Overall, this indicates that the trapped and reflected fractions depend very little on the size or geometry of the field; the main determining factor is the field strength.

\subsection{Dependence on evolutionary stage}\label{sec:dep_profile}
\begin{figure}
  \centering
  \includegraphics[clip=true, trim=0cm 1cm 0cm 0cm, width=0.95\columnwidth]{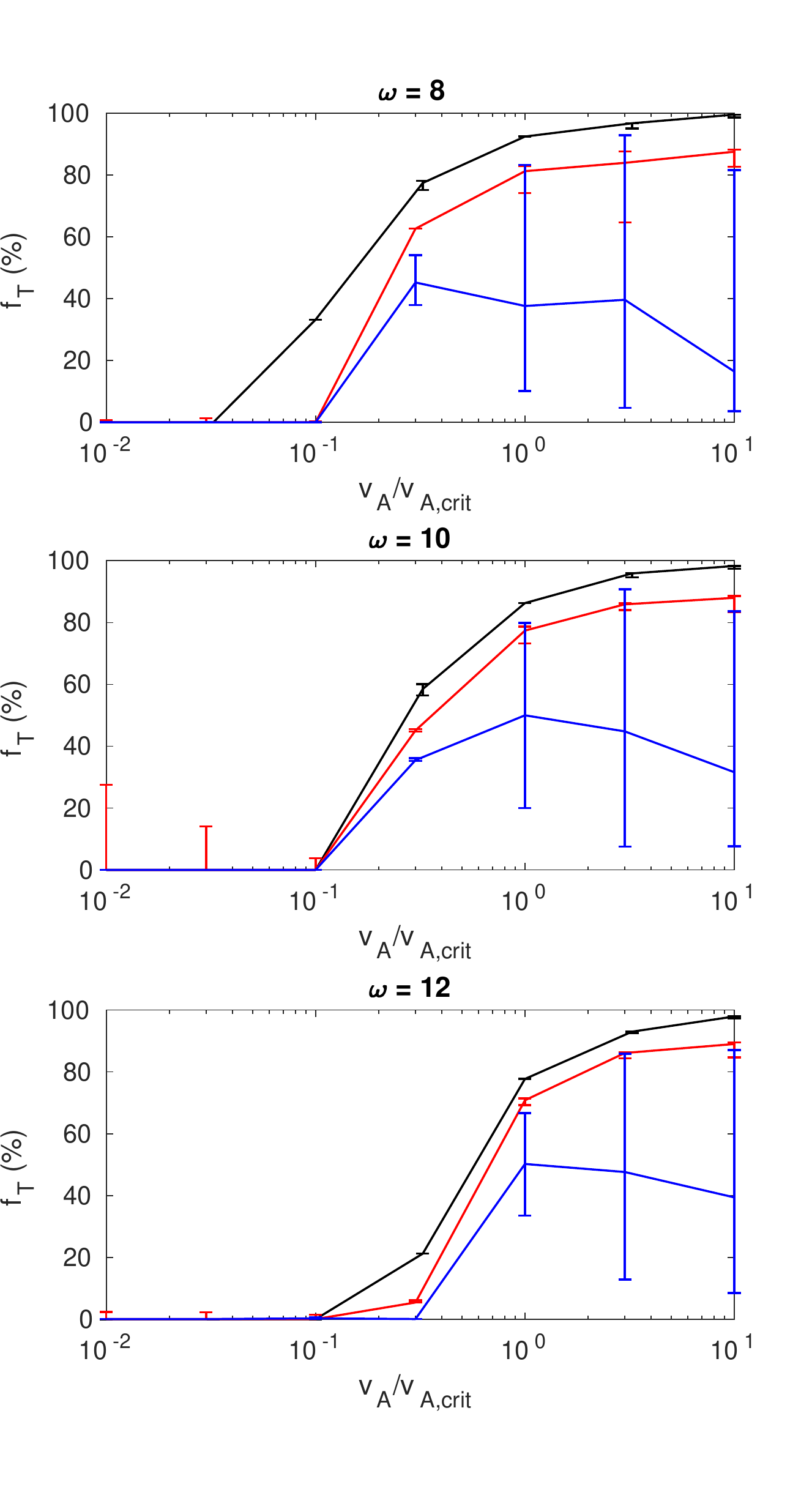}
  \caption{Trapped fraction $f_T$ as a function of field strength, comparing Models A, B, and C, which are shown in black, red and blue, respectively. The field strength is expressed as the ratio between central and critical Alfv\'{e}n speeds. Note that this choice of representation allows for meaningful comparison between the three models, between which $V_{A,\text{crit}}$ differs substantially (see Table \ref{tab:profiles}). The three panels show the results for different wave frequencies, and all results shown are for $\ell = 1$.}
  \label{fig:mgres_profile_om}
\end{figure}

Figure \ref{fig:mgres_profile_om} compares $f_T$ between the three stellar models, these being shown in different colours. While trapping sets in at roughly the same value of $V_{A,\text{cen}}/V_{A,\text{crit}}$ (about 0.1) for all three models, it is interesting to observe that at high field strengths, $f_T$ appears to saturate at a different value depending on the evolutionary state. For the least evolved star (Model A, shown in black) the trapping fraction approaches 100\% at high field strengths, but for Model B (red) this is closer to 80--90\%. For Model C (blue), values between 20--50\% are more characteristic. Despite some large error bars, one can see that at any given ratio of $V_A/V_{A,\text{crit}}$, $f_T$ is systematically lower for a more evolved model.

The explanation behind this is not immediately obvious. A clue as to why more evolved stars might tend to exhibit lower rates of trapping may come from the idealised Cartesian model studied by LP18, which allowed for an analytic ``trapping criterion'' to be derived. We speculate that this may be connected with the increase in the ratio of $N$ to $\omega$ as the star evolves and its core contracts, which affects the chances of satisfying the trapping criterion: see Section \ref{sec:trapping} for further discussion. Tantalisingly, a decrease in $f_T$ as a star evolves is precisely what is required to account for observations, which show decreased dipole mode damping rates with decreasing $\nu_\text{max}$ (more evolved stars have larger $R_*$, and thus lower $\nu_\text{max}$). This is apparent from \citet{Mosser2017a}, figure 7, where the trapping fraction is effectively given by the ratio of black to blue points. 

%% DISCUSSION
\section{Discussion}\label{sec:discussion}
\subsection{Role of critical surfaces}
The concept of a magnetic critical surface was introduced in LP18 and refers to the locus of points where for a given wavevector $\mathbf{k}$, the resonance criterion $\omega_A(\mathbf{k}) = \omega_g(\mathbf{k})$ is satisfied. In that work, which used an idealised Cartesian setup, these were found to be associated with sites where the trapping phenomenon operated. It is of interest to revisit and examine this concept in the more complex, three-dimensional geometry here.

Recall that the resonance criterion was used in the determination of the critical field strength, and so it is when fields reach these strengths that critical surfaces are expected to appear. However, the exact locations of critical surfaces are not straightforward to predict, because $\mathbf{k}$ is not a constant of motion. The expression (\ref{eq:vAcrit}) was derived by assuming that $k_\perp = \sqrt{\ell(\ell+1)}/r$, but this expression only remains true for a ray propagating under conditions of spherical symmetry, which are broken by a magnetic field. Where fields are strong, significant exchanges between $k_r$ (which is generally large to begin with), $k_\theta$ and $k_\phi$ can potentially occur. This means that a ray of initial wavevector $\mathbf{k}_0$, for which one might predict to eventually meet an associated critical surface located where $\omega_g(\mathbf{k}_0) = \omega_A(\mathbf{k}_0)$, could end up evolving its wavevector in a way so as to make that critical surface distort or even disappear before it gets there. Hence to adequately investigate the occurrence and role of critical surfaces, one needs to solve the initial value problem for the ray trajectories.

\begin{figure}
  \centering
  \includegraphics[width=\columnwidth]{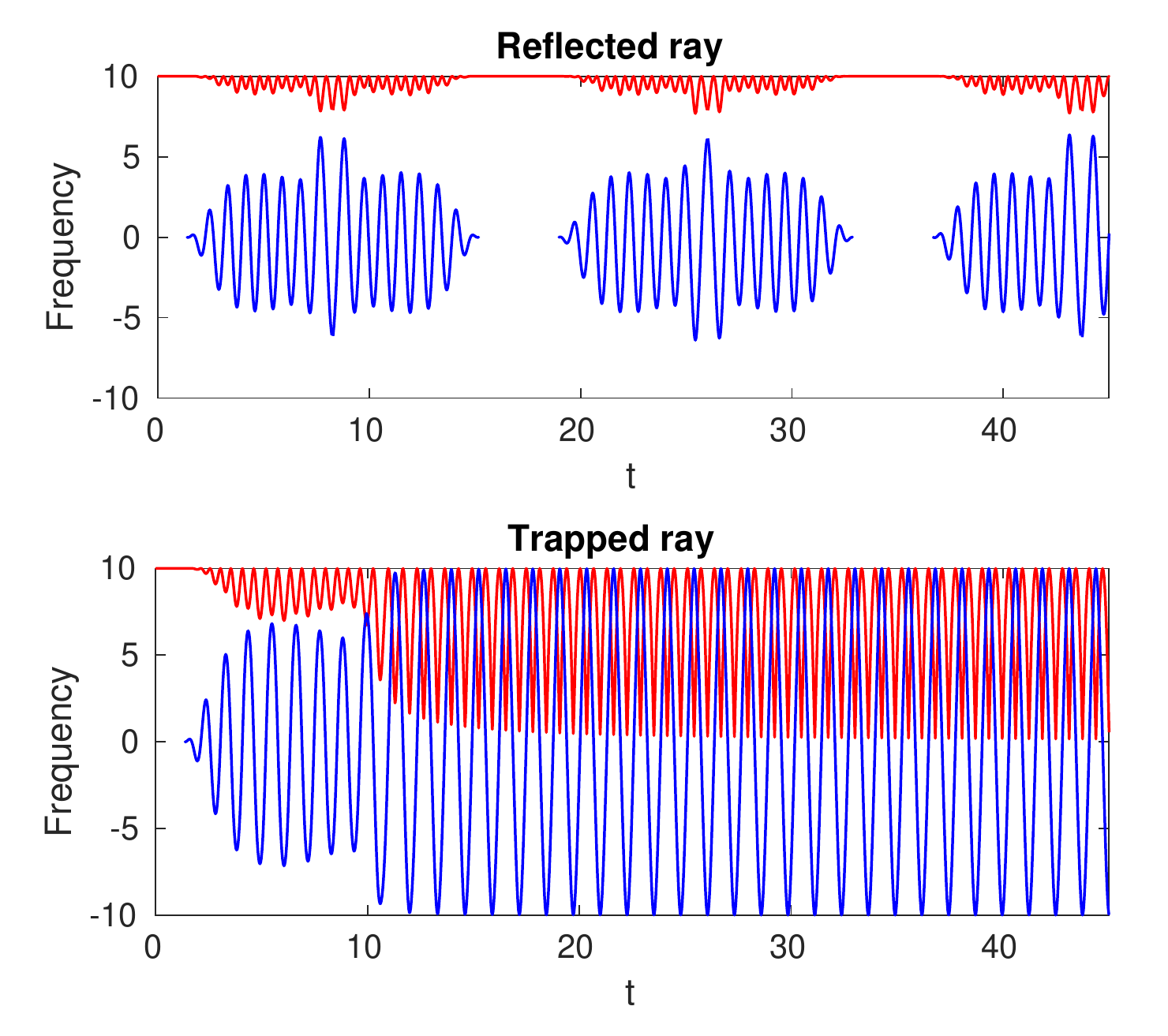}
  \caption{Plots of $\omega_g = \kappa_\perp N$ (red) and $\omega_A = \mathbf{k} \cdot \mathbf{V}_A$ (blue) versus time, for the reflected and trapped rays of Figures \ref{fig:ray137} and \ref{fig:ray204}. In the case of the trapped ray, wavenumber divergence begins at around $t = 10$. Note that the horizontal axis has been truncated prematurely (the total integration was for 100 time units) to enable oscillatory features can be seen more clearly.}
  \label{fig:critical_surfaces}
\end{figure}

Figure \ref{fig:critical_surfaces} plots $\omega_g$ and $\omega_A$ over the course of the trajectory for the reflected and trapped rays of Figures \ref{fig:ray137} and \ref{fig:ray204}. Critical surfaces correspond to where the two curves intersect. In the case of the reflected ray, we can see that $\omega_g > \omega_A$ throughout, whereas for the trapped ray, the red and blue curves begin to intersect around $t = 10$, which coincides with the time at which the wavenumber begins to diverge (Figure \ref{fig:ray204}, bottom). Inspection of these quantities for many other rays suggests that this is a characteristic pattern of behaviour: for trapped rays, the envelopes of $\omega_g$ and $\omega_A$ strongly overlap, with the lower envelope of $\omega_g$ reaching near-zero values. In contrast, for reflected rays the envelopes of $\omega_g$ and $\omega_A$ are largely separate, with the lower envelope of $\omega_g$ usually above or otherwise overlapping minimally with the upper envelope of $\omega_A$. This suggests that reflected rays are those able retain their predominantly gravity-like character through fortuitous paths of propagation.

\begin{figure}
  \centering
  \includegraphics[width=0.95\columnwidth]{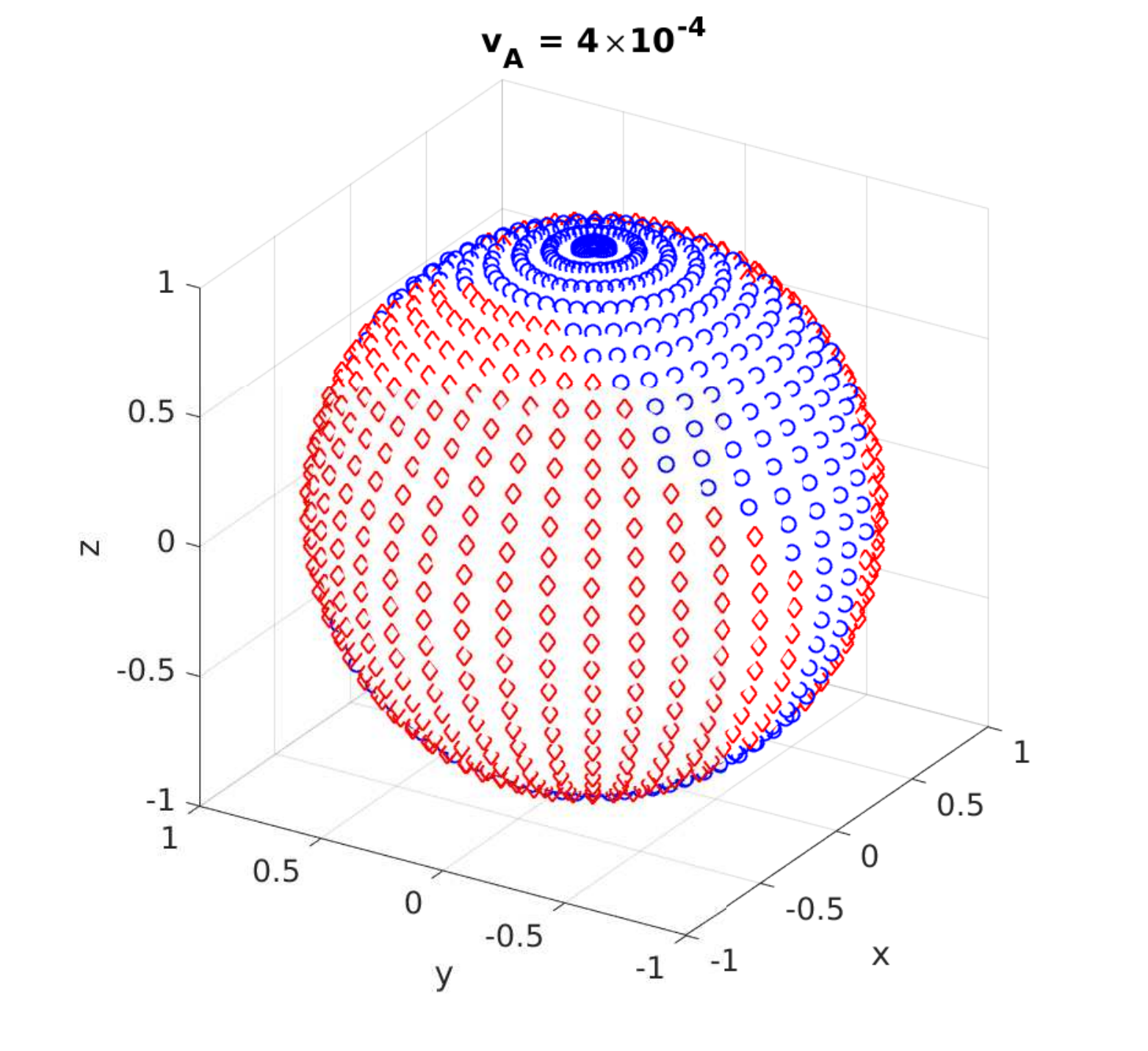}
  \includegraphics[width=0.95\columnwidth]{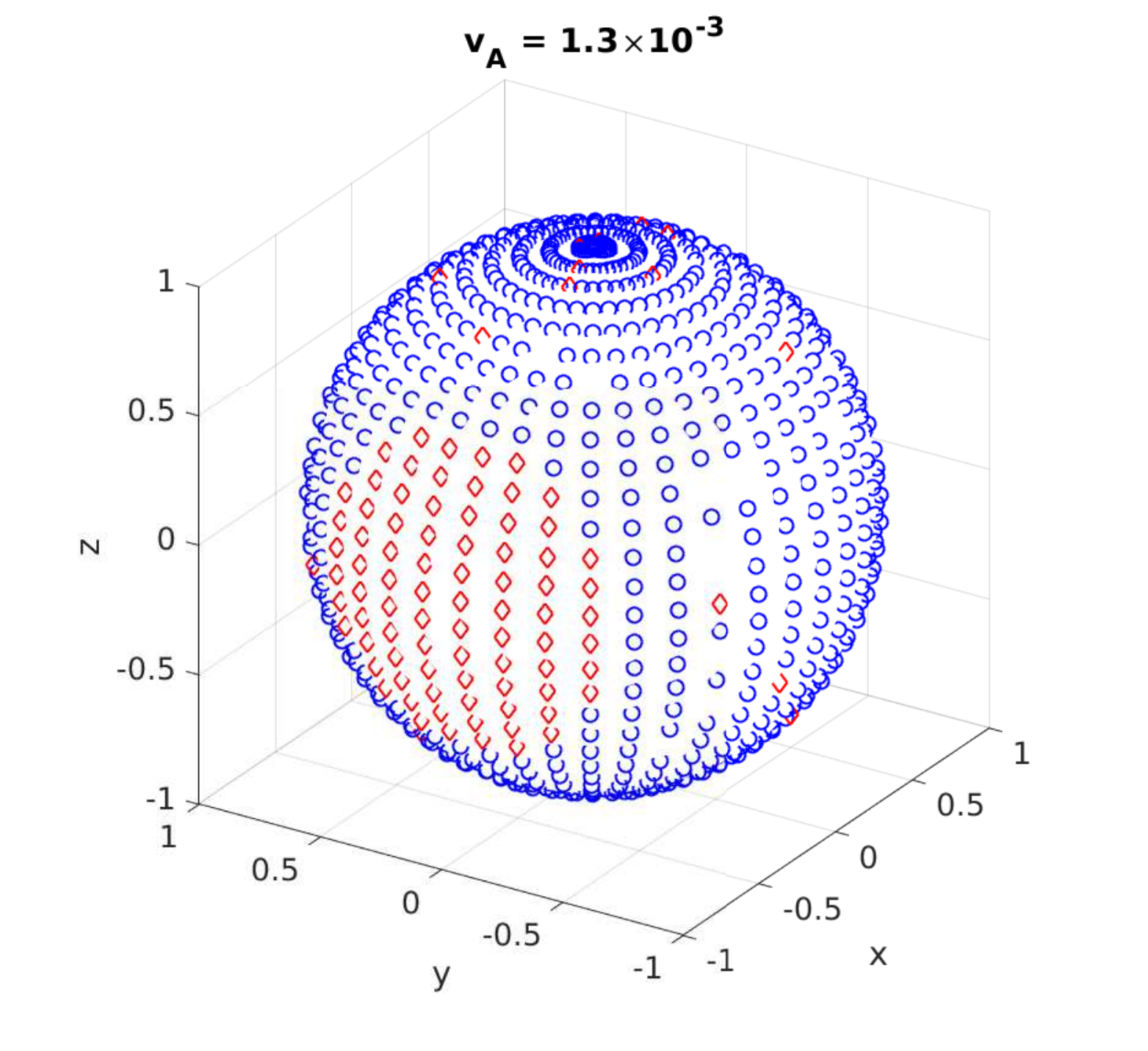}
  \caption{Colour-coded outcome of rays launched with $\omega = 8\,\omega_\text{dyn}$ and $\ell = 1$ into Model A. The field radius was $R_f = 0.016$ and the launch radius was $r_0 = 0.04$; note however that the points have been plotted on a unit sphere, with $(x, y, z) = (\sin \theta_0 \cos \phi_0, \sin \theta_0 \sin \phi_0, \cos \theta_0)$. Red diamonds correspond to reflected rays, while blue circles correspond to trapped rays, and gaps correspond to unclassified or non-convergent rays. Rays were launched along 40 meridians, with 30 rays per meridian, to give a total of 1200 points in each plot. Rays launched on different meridians have different values of $\alpha$: these were initialised such that $\phi_0 = \alpha$. The two panels show the outcomes for two different field strengths, indicated in the header in units of $V_\text{dyn}$. It can be seen that the trapping fraction (area covered by blue points) is larger for the higher field strength. For an impression of the full sphere, refer to the rotating animations in the online supplementary material.}
  \label{fig:trapping_distr}
\end{figure}

As a further remark, we recall that these two rays were launched with identical $k_r$ and $k_\perp$, from the same $r_0$, for the same background stellar model and magnetic field configuration. The fact that one encounters a critical surface and is thereafter trapped, while the other remains free to roam between magnetised and unmagnetised portions of the cavity, strongly underscores the importance of orientation effects (incoming latitude and polarisation) in determining the outcome of a ray. This is illustrated in Figure \ref{fig:trapping_distr}, which plots the angular distributions of trapped and reflected rays, where rays launched along different meridians have different polarisations (given by $\alpha = \phi_0$). The outcome of a ray clearly depends on both quantities, forming interesting segregation patterns in phase space illustrated by the different coloured patches. The main aspect these plots are intended to convey are the fractional areas associated with the two phenomena, which are our desired $f_T$ and $f_R$. The two different panels in Figure \ref{fig:trapping_distr} are for identical starting parameters except for the field strength, demonstrating how the trapping area expands when the field strength is increased.

It can be seen from Figure \ref{fig:trapping_distr} that rays are preferentially reflected if they (i) have been launched from near the magnetic equator, and (ii) have predominantly zonal polarisation vectors, i.e.~aligned/anti-aligned with the $\phi$-direction. This is consistent with the idea that reflected rays are those with small $\omega_A = \mathbf{k} \cdot \mathbf{V}_A$ (never meeting a critical surface): if $\mathbf{k}$ is predominantly radial then this would be true for near-equatorial trajectories, since the field lines of the Prendergast solution are horizontal in the vicinity of the magnetic equator. Note that any field satisfying $\nabla \cdot \mathbf{B} = 0$ must be horizontal somewhere (field lines form closed loops), so reflection should also be possible for other field configurations, including those that do not necessarily have a mixed nature (i.e.~ones that are purely poloidal or purely toroidal). Rather, the key aspect here is the interplay between radial and horizontal field components. The work of \citet{Fuller2015} highlighted the importance of the radial component of the field in interacting with gravity waves; in this work we show that regions where the horizontal component dominates also play an important role, since these allow gravity waves to avoid interaction with the field.

\subsection{Importance of trapping criterion}\label{sec:trapping}
With the idealised geometry used in LP18, it was possible to derive an analytic criterion predicting the onset of trapping given the wave frequency $\omega$ in terms of the angle $\eta_0$ between the magnetic flux surfaces and planes of stratification:
\begin{align}
  \sin \eta_0 < \frac{\omega}{N} \:. \label{eq:trapping_criterion}
\end{align}
One assumption made in deriving this was the lack of variation in any background quantity along magnetic flux surfaces, which was true in the idealised Cartesian setup of LP18 but is not true here. While the above criterion was very successful in predicting the onset of trapping in the earlier work, one might wonder how applicable it remains in the more realistic spherical case.

\begin{figure}
  \centering
  \includegraphics[width=0.95\columnwidth]{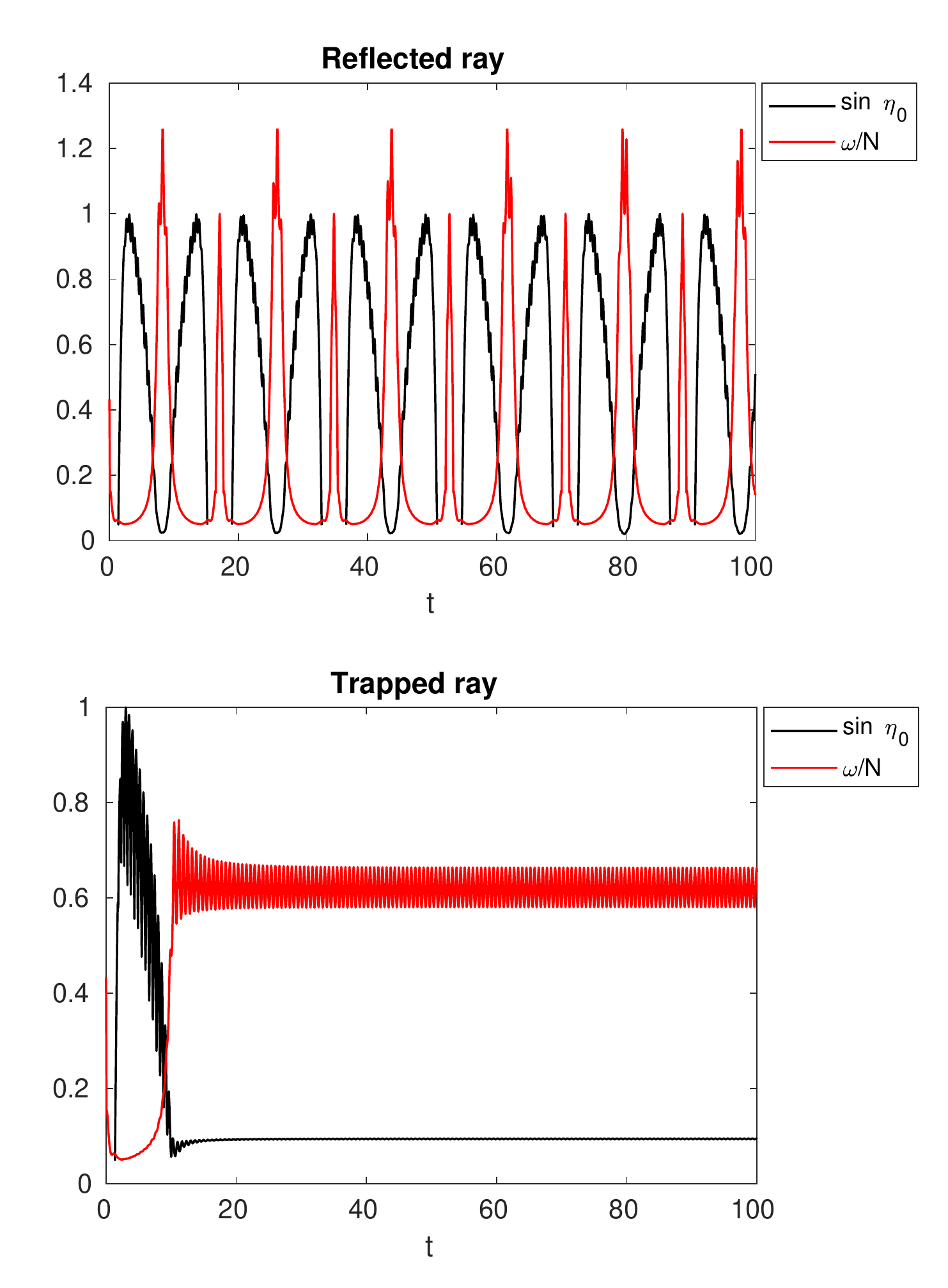}
  \caption{Plots of the LHS (black) and RHS (red) of the trapping criterion (\ref{eq:trapping_criterion}) originally derived by LP18 for an idealised Cartesian setup, for the two rays previously shown in Figures \ref{fig:ray137} and \ref{fig:ray204}. The trapping criterion is nominally satisfied when the black curve falls below the red.}
  \label{fig:trapping_criterion}
\end{figure}

Figure \ref{fig:trapping_criterion} plots the left and right-hand sides of (\ref{eq:trapping_criterion}), for the reflected and trapped ray of Figures \ref{fig:ray137} and \ref{fig:ray204}. In both cases, it can be seen that the trapping criterion is satisfied for a non-zero portion of the trajectory. However for the reflected ray, this only accounts for a small fraction of each orbit, and the ray is always able to subsequently escape. This is not the case for the trapped ray, which may begin with the trapping criterion unsatisfied but is then unable to escape once $\sin \eta_0$ falls below $\omega/N$, which thereafter remains true indefinitely. The point at which this occurs, around $t = 10$ for this particular ray, is observed coincide with the beginning of wavenumber divergence, which also coincides with the point at which $\omega_g$ and $\omega_A$ begin to overlap. Inspection of more rays shows this qualitative behaviour to be characteristic of reflected and trapped rays in general. Thus it appears as though the trapping criterion still carries significance in the full spherical case, but the role it commands is less strict compared to in the idealised Cartesian setup under which it was originally derived.

A further comment relates to the width of the red peaks in Figure \ref{fig:trapping_criterion} (top, corresponding to $\omega/N$ versus $t$) for reflected rays in the more evolved stellar models. As a result of the increasing stratification from ongoing contraction of the core, the widths of the red peaks, which occur in the vicinity of turning points, decrease as the star evolves. Reflected rays in more evolved stars thus spend a smaller fraction of their orbit with $\sin \eta_0 < \omega/N$ satisfied. We speculate that this in turn may be related to the trend for the more evolved stellar models (B and C) to show systematically lower $f_T$, as seen from Figure \ref{fig:mgres_profile_om}: if $\omega/N$ is smaller, then a ray spends less time with $\sin \eta_0 < \omega/N$ satisfied, which may reduce its chances of becoming trapped.

\subsection{Relevance for dipole dichotomy problem}
The source of energy for driving oscillations in evolved stars lies in convective motions occurring in the envelope. These generate acoustic waves that tunnel through an evanescent zone into the stably-stratified core, resuming propagation as gravity waves. In the absence of loss processes, these gravity waves will reflect off the inner turning point of the g-mode cavity and tunnel back out into the envelope. At frequencies for which constructive interference occurs, a global standing wave (mode of oscillation) forms. Now a ubiquitous route of energy loss in solar-like oscillators, for both radial and non-radial modes, is convective damping occurring in the envelope. The time scale $\tau_c$ associated with this process can be measured from radial mode line widths and is characteristically about 15\,d for red giants \citep[][figure 3b]{Mosser2017a}. The dipole dichotomy problem refers to the existence of a group of stars whose dipole (and higher multipole) modes exhibit widths and amplitudes that suggest an additional source of damping besides convection.

If a magnetic field of sufficient strength is present in the core, this creates an avenue for dissipation through the trapping phenomenon, characterised by wavenumber divergence. The time scale associated with this energy loss process is
\begin{align}
  \tau_m = \left( \nu_\text{dyn} T^2 f_T \right)^{-1} \:,
\end{align}
where $1/\nu_\text{dyn}$ is the information crossing time of the star, $T^2$ is the square of the transmission coefficient through the evanescent zone (fraction of acoustic energy transferred into gravity wave motions), and $f_T$ is the fraction of gravity waves trapped and dissipated by the magnetic field. Typical values of $T^2 \sim 0.45$ have been measured for red giants \citep[][figure 3d]{Mosser2017a}.

The quantity $\tau_m$ is closely tied to the quantity $\mathcal{V}$, which is defined \citep[after][therein $v$]{Mosser2017a} to be the ratio of squared visibilities of ``depressed'' to ``normal'' stars. Equivalently, this is the inverse of the ratio of their respective damping rates. Assuming that normal stars experience damping solely through convection, while depressed stars experience damping both through convection and magnetic effects, it follows that
\begin{align}
  \mathcal{V} = \frac{1}{1 + \tau_c/\tau_m} \:.
\end{align}
Using the characteristic values for $T^2$ and $\tau_c$ mentioned above, for $\ell = 1$, $\omega = 10$, $V_{A,\text{cen}} = V_{A,\text{crit}}$ and substituting appropriate values for $\nu_\text{dyn}$ and $f_T$, one finds that $\tau_m = 1.9$, 3.6 and 21\,d for Models A, B and C, respectively. The corresponding values of $\mathcal{V}$ are 0.11, 0.19 and 0.58. This seems to agree well with observations, which show $\mathcal{V}$ increasing with decreasing $\nu_\text{max}$ \citep[cf.][figure 7]{Mosser2017a}.

Our results show that for a given frequency, higher degrees experience larger rates of trapping $f_T$ (see Figures \ref{fig:profile26_mgres_om_l}--\ref{fig:profile30_mgres_Rf_l}), which seemingly contradicts the observation that quadrupole and octupole modes exhibit less depression compared to dipole modes. However, this is resolved if one factors in the $\ell$-dependence of the transmission coefficient
\begin{align}
  T = \exp\left( \int_{r_g}^{r_p} \rmi k_r dr \right) \:, \label{eq:T_coeff}
\end{align}
where $k_r$ may be given approximately by
\begin{align}
  k_r^2 \approx \frac{(S_\ell^2 - \omega^2) (N^2 - \omega^2)}{\omega^2 c_s^2} \:.
\end{align}
Here $r_g$ and $r_p$ are the radial coordinates of the upper and lower turning points of the g- and p-mode cavities respectively, $c_s$ is the sound speed, and $S_\ell = c_s \sqrt{\ell(\ell+1)} / r$ is the Lamb frequency. Now the observable quantity is the visibility $\mathcal{V}$, which depends on both $T$ and $f_T$ through $\tau_m$. Estimation of $T^2$ values for Model B using the above expression (\ref{eq:T_coeff}) yields 0.4, 0.03 and $10^{-3}$ for $\ell = 1$, 2 and 3, where we take $\omega = 10\,\omega_\text{dyn}$. At $V_A = 0.3\,V_{A,\text{crit}}$, $f_T$ values are approximately 0.55, 0.65 and 0.7 for $\ell = 1$, 2 and 3, so then correspondingly $T^2 f_T = 0.2$, 0.02 and $7 \times 10^{-4}$. We see that the increase in damping time $\tau_m$ with $\ell$ is thus driven by the steep decrease in $T^2$, offsetting the increase in $f_T$. With these estimates for $T^2$, the visibilities for $\ell=2$ and $\ell=3$ (defined likewise to be normalised with respect to ``normal'' quadrupole and octupole modes) are then $\mathcal{V} = 0.83$ and 0.99, respectively. This is quantitatively consistent with the observational data, which show mild depression for quadrupole modes and negligible depression for octupole modes; cf.~figure 5 of \citet{Stello2016a}, where $\mathcal{V}$ corresponds to the ratio of solid to open markers.

The primary difference with the previous work of \citet{Fuller2015} considering the potential damping role of magnetic fields is the introduction of the factor $f_T$, which we have shown can be substantially less than unity. It has been established that if this is unity (magnetic fields dissipate 100\% of wave energy in the core), then the observed values of $\mathcal{V}$ cannot be reproduced \citep{Mosser2017a, Arentoft2017}. This has been used as an argument against a magnetic explanation for the dipole dichotomy problem. However, our results demonstrate that if orientation effects are taken properly into account, then even at field strengths at or exceeding the critical value, only a fraction of incoming gravity waves will undergo trapping and dissipation by the magnetic field. The rest behave as waves which are dominantly gravity-like in character ($\omega_g > \omega_A\,\forall t$) and these survive their passage into and out of the magnetised core.

Our measurements of $f_T$ as a function of $\omega$ succeed in reproducing the positive gradient of $\mathcal{V}$ with respect to $\omega$ seen in several red giants (see Section \ref{sec:dep_om}). It appears that this is expected to occur for field strengths in the vicinity of $V_{A,\text{crit}}$ itself or a factor of a few lower. Notice from Table \ref{tab:profiles} that the critical field strength decreases as a star ages. Physically, this is driven by an increase in $N$ and decrease in $\omega$, which play a role in the resonance criterion (\ref{eq:vAcrit}). If magnetic flux is conserved, then field strengths would increase as stars evolve and their cores contract. It may be that the youngest RGB stars are just beginning to venture into the supercritical regime, and the ones with observed gradients in $\mathcal{V}$ with respect to $\omega$ are those caught undergoing this transition. Further along on the RGB, the decrease in $V_{A,\text{crit}}$ and increase in $V_{A,\text{cen}}$ with age ensures that such stars would remain in the supercritical regime. Interestingly, $f_T$ is generally not observed to reach 100\% for increasing field strengths, but appears to plateau at a lower value for older stars. We propose that this may be to do with the decrease in $\omega/N$ with age, since this affects the chances of a random ingoing wave satisfying the trapping criterion (see discussion in Section \ref{sec:trapping}).

To conclude this section, our results show that observations of less than 100\% loss of g-mode energy can be consistent with the possibility that strong magnetic fields exist inside the cores of evolved stars. Gravity wave propagation is possible even when strong fields are present: we have directly shown for realistic stellar models and field configurations that propagation paths exist where the waves remained restored predominantly by buoyancy. It is plausible that the seismic properties (such as asymptotic period spacings) of modes formed by wave propagation along such paths may not differ measurably between depressed and normal stars, although this has yet to be rigorously shown. Also, we make no attempt to rule out other possible mechanisms for damping localised to the core: for example, there are suggestions that this may alternatively be achievable through non-linear wave breaking \citep{Weinberg2019}.

\subsection{Limitations}
While having the advantage of computational efficiency over direct numerical simulations, there are certain aspects of wave propagation that the Hamiltonian ray tracing technique used here cannot capture, because it only incorporates WKBJ to zeroth order. These include evanescence and tunnelling, partial transmission, and amplitude growth/decay. However, none of the above interfere with the objective of this work, which was to measure the reflected/trapped fractions, since reflection and trapping are processes that occur in the regime of full propagation. It remains for the results of this work to be compared and verified with full numerical simulations, which would also be able to take a larger range of physical processes into account.

Rotation has not been incorporated, although we remark that the typically small rotation frequencies (compared with wave frequencies) in evolved stars mean that resonant interactions of the type examined here are unlikely to occur between inertial and acoustic/gravity waves. We therefore do not anticipate that the inclusion of rotation will have a significant impact on our results, although this has yet to be more carefully investigated.

%% SUMMARY
\section{Summary}\label{sec:summary}
We have performed a Hamiltonian ray tracing study of magneto-gravity wave packet dynamics using realistic stellar models and magnetic field configurations, and measured the rates of trapping and reflection ($f_T$ and $f_R = 1 - f_T$) as a function of various parameters. We have found that:
\begin{itemize}
\item There exist trajectories in a strongly magnetised core where propagation is still restored mainly by buoyancy, and such waves experience reflection from the inner turning point of the g-mode cavity much like pure gravity waves;
\item The remaining trajectories correspond to trapped waves, possessing both significant Alfv\'{e}n and gravity wave character, and their wavenumbers are divergent implying eventual dissipation;
\item The onset of trapping occurs when field strengths approach the critical magnetic field strength $B_\text{crit} = V_{A,\text{crit}} \sqrt{\mu_0 \rho}$, where $V_{A,\text{crit}}$ is given in (\ref{eq:vAcrit});
\item As field strengths approach critical values, $f_T$ does not jump from zero to unity but rather increases gradually from zero to a value generally less than unity;
\item The limiting value of $f_T$ is smaller for more evolved stars;
\item The radial extent and detailed geometry of the magnetic field have negligible effect on $f_T$;
\item The outcome of a ray (trapping or reflection) depends crucially on its launch colatitude $\theta_0$ and polarisation $\alpha = \tan^{-1} (k_\theta/k_\phi)$, underscoring the importance of orientation effects.
\end{itemize}
Limitations of our method include the assumption of short wavelengths compared to scales of background variation, the inability to capture higher-order WKBJ effects such as wave amplitude growth/decay, and neglect of rotation. However, for reasons previously discussed, these are not expected to affect the main conclusions of this work.

In broader context, these results would seem to resolve part of the controversy regarding the viability of a magnetic explanation for the dipole dichotomy problem, since they demonstrate that strong fields lead only to partial g-mode energy loss when orientation effects are accounted for. The manner in which the trapping varies with mode frequency and evolutionary stage also appears to be consistent with observations. However, there still remain unanswered questions about the magnetic hypothesis, such as the effect on asymptotic period spacings, which represents another observational constraint to be addressed.

\section*{Acknowledgements}
We thank John Papaloizou for useful feedback, and the referee, Beno\^{i}t Mosser, for thoughtful comments that significantly helped to tighten the links between theoretical and observational results. STL is supported by funding from Churchill College, Cambridge, through a Junior Research Fellowship. 

%%%%%%%%%%%%%%%%%%%%%%%%%%%%%%%%%%%%%%%%%%%%%%%%%%



\bibliographystyle{mnras}
%\bibliography{C:/Users/STCLoi/Documents/articles_and_papers/my_stuff/refs}

%%%%%%%%%%%%%%%%%%%%%%%%%%%%%%%%%%%%%%%%%%%%%%%%%%

%%%%%%%%%%%%%%%%% APPENDICES %%%%%%%%%%%%%%%%%%%%%

\appendix

\section{MESA inlist}\label{sec:inlist}
The 2\,M$_\odot$ model, out of which the three evolutionary snapshots were extracted, was generated by MESA using the following inlist:
\begin{verbatim}
&star_job

  ! begin with a pre-main sequence model
    create_pre_main_sequence_model = .true.

  ! save a model at the end of the run
    save_model_when_terminate = .true.
    save_model_filename = `2M_RG_hr100.mod'

  ! display on-screen plots
    pgstar_flag = .true.

/ !end of star_job namelist

&controls

  ! starting specifications
    initial_mass = 2 ! in Msun units

  ! stopping condition
    max_age = 1.076d9

  ! mesh adjustment
    mesh_delta_coeff = 0.2d0
    max_dq = 1d-3

/ ! end of controls namelist
\end{verbatim}

\section{Ray tracing equations}\label{sec:rt_derivation}
In this section we show how the ray-tracing equations for a general time-independent Hamiltonian $\omega = \omega(\mathbf{x}, \mathbf{k})$ in spherical polar coordinates can be derived. The first of (\ref{eq:Hamilton}) is straightforward: we have
\begin{align}
  \frac{\rmd \mathbf{x}}{\rmd t} &= \left( \frac{\rmd r}{\rmd t},\; r \frac{\rmd \theta}{\rmd t},\; r \sin \theta \frac{\rmd \phi}{\rmd t} \right) \:, \\
  \shortintertext{which, equating with the RHS, leads to}
  \frac{\rmd r}{\rmd t} &= \frac{\del \omega}{\del k_r} \:, \label{eq:drdt_gen} \\
  \frac{\rmd \theta}{\rmd t} &= \frac{1}{r} \frac{\del \omega}{\del k_\theta} \:, \label{eq:dthetadt_gen} \\
  \frac{\rmd \phi}{\rmd t} &= \frac{1}{r \sin \theta} \frac{\del \omega}{\del k_\phi} \:. \label{eq:dphidt_gen}
\end{align}

The second of (\ref{eq:Hamilton}) is more involved, since spatial variation of the unit vectors must be accounted for, e.g.
\begin{align}
  \frac{\rmd k_r}{\rmd t} = \frac{\rmd \mathbf{k}}{\rmd t} \cdot \hat{\mathbf{r}} + \mathbf{k} \cdot \frac{\rmd \hat{\mathbf{r}}}{\rmd t} \:. 
\end{align}
Consider that
\begin{align}
  \frac{\rmd \mathbf{k}}{\rmd t} &= -\frac{\del \omega}{\del r} \nabla r - \frac{\del \omega}{\del \theta} \nabla \theta - \frac{\del \omega}{\del \phi} \nabla \phi - \frac{\del \omega}{\del k_r} (\nabla \hat{\mathbf{r}}) \cdot \mathbf{k} \nonumber \\
  &\qquad - \frac{\del \omega}{\del k_\theta} (\nabla \hat{\boldsymbol{\theta}}) \cdot \mathbf{k} - \frac{\del \omega}{\del k_\phi} (\nabla \hat{\boldsymbol{\phi}}) \cdot \mathbf{k} \\
  \shortintertext{and}
  \frac{\rmd \hat{\mathbf{r}}}{\rmd t} &= \frac{\rmd \mathbf{x}}{\rmd t} \cdot \nabla \hat{\mathbf{r}} = \left( \frac{\del \omega}{\del k_r} \hat{\mathbf{r}} + \frac{\del \omega}{\del k_\theta} \hat{\boldsymbol{\theta}} + \frac{\del \omega}{\del k_\phi} \hat{\boldsymbol{\phi}} \right) \cdot \nabla \hat{\mathbf{r}} \:.
\end{align}
These combine to yield
\begin{align}
  \frac{\rmd k_r}{\rmd t} &= \frac{\del \omega}{\del k_\theta} \left( \hat{\boldsymbol{\theta}} \cdot \nabla \hat{\mathbf{r}} - \hat{\mathbf{r}} \cdot \nabla \hat{\boldsymbol{\theta}} \right) \cdot \mathbf{k} + \frac{\del \omega}{\del k_\phi} \left( \hat{\boldsymbol{\phi}} \cdot \nabla \hat{\mathbf{r}} - \hat{\mathbf{r}} \cdot \nabla \hat{\boldsymbol{\phi}} \right) \cdot \mathbf{k} \nonumber \\
  &\qquad - \frac{\del \omega}{\del r} \hat{\mathbf{r}} \cdot \nabla r - \frac{\del \omega}{\del \theta} \hat{\mathbf{r}} \cdot \nabla \theta - \frac{\del \omega}{\del \phi} \hat{\mathbf{r}} \cdot \nabla \phi \:.
\end{align}
Identical manipulations for the other two components yields
\begin{align}
  \frac{\rmd k_\theta}{\rmd t} &= \frac{\del \omega}{\del k_r} \left( \hat{\mathbf{r}} \cdot \nabla \hat{\boldsymbol{\theta}} - \hat{\boldsymbol{\theta}} \cdot \nabla \hat{\mathbf{r}} \right) \cdot \mathbf{k} + \frac{\del \omega}{\del k_\phi} \left( \hat{\boldsymbol{\phi}} \cdot \nabla \hat{\boldsymbol{\theta}} - \hat{\boldsymbol{\theta}} \cdot \nabla \hat{\boldsymbol{\phi}} \right) \cdot \mathbf{k} \nonumber \\
  &\qquad - \frac{\del \omega}{\del r} \hat{\boldsymbol{\theta}} \cdot \nabla r - \frac{\del \omega}{\del \theta} \hat{\boldsymbol{\theta}} \cdot \nabla \theta - \frac{\del \omega}{\del \phi} \hat{\boldsymbol{\theta}} \cdot \nabla \phi \:, \\
  \frac{\rmd k_\phi}{\rmd t} &= \frac{\del \omega}{\del k_r} \left( \hat{\mathbf{r}} \cdot \nabla \hat{\boldsymbol{\phi}} - \hat{\boldsymbol{\phi}} \cdot \nabla \hat{\mathbf{r}} \right) \cdot \mathbf{k} + \frac{\del \omega}{\del k_\theta} \left( \hat{\boldsymbol{\theta}} \cdot \nabla \hat{\boldsymbol{\phi}} - \hat{\boldsymbol{\phi}} \cdot \nabla \hat{\boldsymbol{\theta}} \right) \cdot \mathbf{k} \nonumber \\
  &\qquad - \frac{\del \omega}{\del r} \hat{\boldsymbol{\phi}} \cdot \nabla r - \frac{\del \omega}{\del \theta} \hat{\boldsymbol{\phi}} \cdot \nabla \theta - \frac{\del \omega}{\del \phi} \hat{\boldsymbol{\phi}} \cdot \nabla \phi \:.
\end{align}
Using
\begin{align}
  \nabla r &= \hat{\mathbf{r}} \:, \\
  \nabla \theta &= \frac{1}{r} \hat{\boldsymbol{\theta}} \:, \\
  \nabla \phi &= \frac{1}{r \sin \theta} \hat{\boldsymbol{\phi}} \:, \\
  \nabla \hat{\mathbf{r}} &= \frac{1}{r} \left( \hat{\boldsymbol{\theta}} \hat{\boldsymbol{\theta}}^\top + \hat{\boldsymbol{\phi}} \hat{\boldsymbol{\phi}}^\top \right) \:, \\
  \nabla \hat{\boldsymbol{\theta}} &= \frac{\cot \theta}{r} \hat{\boldsymbol{\phi}} \hat{\boldsymbol{\phi}}^\top - \frac{1}{r} \hat{\boldsymbol{\theta}} \hat{\mathbf{r}}^\top \:, \\
  \nabla \hat{\boldsymbol{\phi}} &= -\frac{\cot \theta}{r} \hat{\boldsymbol{\phi}} \hat{\boldsymbol{\theta}}^\top - \frac{1}{r} \hat{\boldsymbol{\phi}} \hat{\mathbf{r}}^\top \:,
\end{align}
we arrive at
\begin{align}
  \frac{\rmd k_r}{\rmd t} &= \frac{k_\theta}{r} \frac{\del \omega}{\del k_\theta} + \frac{k_\phi}{r} \frac{\del \omega}{\del k_\phi} - \frac{\del \omega}{\del r} \:, \label{eq:dkrdt_gen} \\
  \frac{\rmd k_\theta}{\rmd t} &= -\frac{k_\theta}{r} \frac{\del \omega}{\del k_r} + \frac{k_\phi \cot \theta}{r} \frac{\del \omega}{\del k_\phi} - \frac{1}{r} \frac{\del \omega}{\del \theta} \:, \label{eq:dkthetadt_gen} \\
  \frac{\rmd k_\phi}{\rmd t} &= -\frac{k_\phi}{r} \frac{\del \omega}{\del k_r} - \frac{k_\phi \cot \theta}{r} \frac{\del \omega}{\del k_\theta} - \frac{1}{r \sin \theta} \frac{\del \omega}{\del \phi} \:. \label{eq:dkphidt_gen}
\end{align}
Equations (\ref{eq:drdt_gen})--(\ref{eq:dphidt_gen}) and (\ref{eq:dkrdt_gen})--(\ref{eq:dkphidt_gen}) are the ray tracing equations for a general time-independent Hamiltonian with no assumptions about symmetry.

Substitution of (\ref{eq:MG_DR}) into (\ref{eq:drdt_gen})--(\ref{eq:dphidt_gen}), (\ref{eq:dkrdt_gen})--(\ref{eq:dkphidt_gen}), along with the assumptions that $\del/\del \phi \equiv 0$ and $N = N(r)$, yields Equations (\ref{eq:drdt})--(\ref{eq:dkphidt}) which are for magneto-gravity waves.

\section{Conserving wave frequency}\label{sec:omega_conservation}
The RK4 scheme used does not inherently conserve $\omega$. Prior to implementation of the routine described below, $\omega$ was often seen to drift over the course of a trajectory as a result of accumulated truncation errors. To enforce conservation of $\omega$ for each ray, we chose to adjust the value of $k_r$ at the end of each time step in a way that would bring $\omega$ back to its initial value. The magneto-gravity dispersion relation (\ref{eq:MG_DR}) can be rewritten as a quartic in $k_r$:
\begin{align}
  0 &= a_1 k_r^4 + a_2 k_r^3 + a_3 k_r^2 + a_4 k_r + a_5 \:, \label{eq:kr_quartic} \\
  \shortintertext{where}
  a_1 &= V_{Ar}^2 \:, \\
  a_2 &= 2V_{Ar} \left( k_\theta V_{A\theta} + k_\phi V_{A\phi} \right) \:, \\
  a_3 &= V_{Ar}^2 k_\perp^2 + (k_\theta V_{A\theta} + k_\phi V_{A\phi})^2 - \omega^2 \:, \\
  a_4 &= 2V_{Ar} k_\perp^2 (k_\theta V_{A\theta} + k_\phi V_{A\phi}) \:, \\
  a_5 &= k_\perp^2 \left[ (k_\theta V_{A\theta} + k_\phi V_{A\phi})^2 + N^2 - \omega^2 \right] \:.
\end{align}
In practice, truncation errors meant that the RHS of (\ref{eq:kr_quartic}) was not always exactly zero after each time step. The goal of the correction procedure was to find the (likely nearby) value of $k_r$ satisfying (\ref{eq:kr_quartic}). The old $k_r$ value would be replaced by this value for the next time step.

To solve (\ref{eq:kr_quartic}), Newton-Raphson iteration was performed for a maximum of 5 iterations or until successive changes in $k_r$ were less than 1\%, for each ray. The most recent value of $k_r$ calculated was used to begin the iteration; this was typically close enough to the nearest root of (\ref{eq:kr_quartic}) for convergence to occur within about 3 iterations. However, rays would occasionally fail to converge, which could be detected through the resulting fluctuations in $\omega$. Such rays were discarded (more detail in Section \ref{sec:quality}). 

We remark that alternative attempts to use a symplectic method, namely 4th-order Gauss-Legendre Runge-Kutta (which inherently conserves the Hamiltonian), were also fraught with convergence difficulties, in this case associated with implicit intermediate steps. The explicit RK4 scheme combined with a post-correction step was found to behave better, and so we opted for this over the symplectic method.

%%%%%%%%%%%%%%%%%%%%%%%%%%%%%%%%%%%%%%%%%%%%%%%%%%

% Don't change these lines
\bsp	% typesetting comment
\label{lastpage}
\end{document}